\documentclass[aps,amsmath,prd,reprint,showpacs,preprintnumbers,twocolumn]{revtex4}

\usepackage{color}
\usepackage{graphicx}
\usepackage{latexsym}
\usepackage{bm}
\usepackage{amsmath}
\usepackage{slashed}
\usepackage{amsfonts}

\def\lprt{\stackrel{\hspace{.45em}\reflectbox{\scalebox{1}[-1]{\raisebox{-0.49em}[0.2em][0em]{$\vec{}$}}}}{\raisebox{0em}{$\partial$}}}

\def\lnabla{\stackrel{\hspace{.25em}\reflectbox{\scalebox{1}[-1]{\raisebox{-0.49em}[0.2em][0em]{$\vec{}$}}}}{\raisebox{0em}{$\nabla$}}}

\def\lsquare{\stackrel{\hspace{.25em}\reflectbox{\scalebox{1}[-1]{\raisebox{-0.49em}[0.2em][0em]{$\vec{}$}}}}{\raisebox{0em}{$\square$}}}

\def\half{{\textstyle\frac12}}
\def\quar{{\textstyle\frac14}}

\def\lsim{\mathrel{\rlap{\lower3pt\hbox{$\sim$}}
    \raise2pt\hbox{$<$}}}
\def\gsim{\mathrel{\rlap{\lower3pt\hbox{$\sim$}}
    \raise2pt\hbox{$>$}}}
\def\sqr#1#2{{\vcenter{\vbox{\hrule height.#2pt
         \hbox{\vrule width.#2pt height#1pt \kern#1pt
         \vrule width.#2pt}
         \hrule height.#2pt}}}}

\def\prt{\partial}
\def\lrpartial{\raise 1pt\hbox{$\stackrel\leftrightarrow\partial$}}

\newcommand{\rf}[1]{(\ref{#1})}

\def\etal{{\it et al.}}

\begin{document}

\title{Asymptotic states and renormalization
in Lorentz-violating quantum field theory}

\author{Mauro Cambiaso}
\affiliation{Universidad Andres Bello, Departamento de Ciencias Fisicas,
Facultad de Ciencias Exactas, Avenida Republica 220, Santiago, Chile}

\author{Ralf Lehnert}
\affiliation{Indiana University Center for Spacetime Symmetries, 
Bloomington, Indiana 47405, USA}

\author{Robertus Potting}
\affiliation{CENTRA and Departamento de F\'\i sica, FCT, Universidade do Algarve, 8005-139 Faro, Portugal}

\date{September 9, 2014}

\begin{abstract} 
Asymptotic single-particle states in quantum field theories 
with small departures from Lorentz symmetry 
are investigated 
perturbatively
with focus on potential phenomenological ramifications.
To this end, 
one-loop radiative corrections for a sample Lorentz-violating Lagrangian 
contained in the Standard-Model Extension (SME) 
are studied at linear order in Lorentz breakdown. 
It is found that
the spinor kinetic operator,
and thus the free-particle physics, 
is modified by
Lorentz-violating operators 
absent from the original Lagrangian.
As a consequence of this result,
both the standard renormalization procedure 
as well as the Lehmann--Symanzik--Zimmermann reduction formalism
need to be adapted. 
The necessary adaptations  
are worked out explicitly at first order in Lorentz-breaking coefficients.
\end{abstract}


\maketitle

\section{Introduction}
\label{introduction}

Current understanding of physics 
at the fundamental level 
is based on two distinct theories: 
general relativity (GR) and the Standard Model (SM) of particle physics.
It is commonly believed that 
these two theories arise as the low-energy limit 
of an underlying Planck-scale framework that 
consistently merges gravity and quantum mechanics.  
Since direct measurements at this scale are presently impractical, 
experimental research in this field relies largely on 
ultrahigh-precision searches for Planck-suppressed effects 
at attainable energies.

One possible effect in this context is a minute breakdown of
Lorentz invariance.
Lorentz symmetry is a fundamental feature of both GR and the SM, 
so that any observed deviation from this symmetry 
would imply new physics.
A number of theoretical approaches to physics beyond the SM, 
such as strings~\cite{ksp}, 
noncommutative field theories~\cite{ncqed}, 
cosmologically varying scalar fields~\cite{spacetimevarying}, 
quantum gravity~\cite{qg}, 
random-dynamics models~\cite{fn}, 
multiverses~\cite{bj}, 
brane-world scenarios~\cite{brane}, 
and massive gravity~\cite{modgrav}, 
are believed to allow for small violations of Lorentz invariance 
at low energies.
Searches for such violations are also motivated by 
the apparent fundamental character of Lorentz symmetry.
Consequently, 
Lorentz invariance ought to be supported as firmly as possible 
by experimental evidence.

It is natural to expect that 
Lorentz-violating effects can be described within effective field theory, 
at least at currently attainable energies~\cite{kp}.
The framework generally adopted in this context
is the Standard-Model Extension (SME)~\cite{sme,ssb},
which contains both GR and the SM 
as limiting cases.
The additional Lagrangian terms present in the SME 
include all operators for Lorentz violation that 
are scalars under coordinate changes.
The SME has constituted the basis 
for the analysis of numerous experimental searches for Lorentz breakdown~\cite{DataTables}.

Paralleling the conventional Lorentz-symmetric case,
perturbative quantum-field analyses within the SME also rely on 
a few key theoretical concepts. 
Some of these, 
such as canonical quantization~\cite{quant1,quant2} and renormalization~\cite{Kostelecky:2001jc,renorm,Shapiro},
have previously been studied and generalized to the SME. 
Another such core concept concerns the treatment of external states.
They span the asymptotic Hilbert space,
so their determination is of fundamental importance 
for perturbation theory.
For example, 
explicit S-matrix calculations 
require a separate, independent determination 
of the external legs up to the desired order. 
This special status of external-leg physics 
is highlighted by the usual Feynman rules:
external-leg corrections cannot be incorporated 
into the diagram for a scattering process; 
the rules for S-matrix calculations specifically call for ``amputed" diagrams. 
The usual treatment of radiative corrections to external legs 
involves sophisticated theoretical concepts like 
the K\"all\'en--Lehmann representation~\cite{KallenLehmann} and 
the Lehmann--Symanzik--Zimmermann (LSZ) reduction formalism~\cite{LSZ}.
Although a number of prior investigations 
have considered radiative corrections
from various other perspectives~\citep{CS,loops},
we are unaware of any dedicated study to generalize 
the Lorentz-invariant external-leg treatment 
to Lorentz-violating field theories.

A second need 
for a proper understanding of the asymptotic Hilbert space 
in the presence of Lorentz breakdown
derives from its phenomenological importance:
external-state effects govern the physics of free particles 
and are therefore also crucial for numerous Lorentz tests.
Examples include 
various kinematical threshold effects in cosmic rays~\cite{crays},
photon birefringence and dispersion~\cite{KM02,cphotons},
collider kinematics and interferometry~\cite{colliders,hohensee,synchrotron,graal},
and neutrino propagation~\cite{neutrinos}.
Paralleling the conventional Lorentz-symmetric case,
all previous analyses have been performed under the tacit assumption that
the physics of free particles is determined by
the quadratic pieces of the corresponding Lagrangian
\cite{Colladay:2001wk}.
However, 
this approach disregards the self-interactions of the particle, 
although such effects are always present, 
even for asymptotic states.
Consequently, 
they need to be considered, 
e.g., 
in any scattering process beyond tree level.
In a conventional renormalizable quantum field theory (QFT), 
Lorentz symmetry implies that 
the quadratic Lagrangian 
can only acquire a mass shift and a field-strength factor, 
both of which can be treated by
renormalization of existing quantities. 
The external QFT legs are then identical in structure 
to the quadratic-Lagrangian solutions, 
which therefore indeed describe the propagation of free particles correctly.
A nonperturbative rigorous justification for this feature 
is given by the aforementioned
LSZ reduction formalism~\cite{LSZ}. 
However, 
in the presence of Lorentz violation 
a similar line of reasoning fails,
and the question regarding the determination of free-particle properties arises.

The present work is intended 
to initiate a theoretical investigation of these issues.  
In particular, 
we demonstrate that 
in the absence of Lorentz symmetry
the external legs in perturbative quantum field theory
exhibit a different structure than
the plane-wave solutions arising from the quadratic Lagrangian~\cite{GravCase}. 
This result is in accordance with a recent work~\cite{Potting:2011yj}
in which a generalization of the K\"all\'en--Lehmann
representation for the propagator
was derived for a field-theoretic model with fermions that are coupled
to the same Lorentz-violating SME coefficients as the ones
we are considering in this work.
In fact, 
we will use the results obtained in Ref.~\cite{Potting:2011yj}
to extract consistently the one-particle poles 
at first order in Lorentz violation.
These poles define the external states of scattering amplitudes.
To this end,
we generalize the conventional LSZ reduction formula to 
include  Lorentz-breaking SME corrections at linear order.

For this analysis,
we restrict ourselves to a subset of the minimal SME's electrodynamics sector
for simplicity.
Moreover, 
we are primarily focused on effects that
may potentially be of phenomenological relevance 
and affect the usual perturbative expansion of quantum field theory.
Our analysis  
is therefore performed at first order in SME coefficients, 
an approach justified on observational grounds.
A future nonperturbative treatment of these issues
within formal field theory 
would be interesting,
but lies outside our present scope.
Throughout, 
we adopt natural units $c=\hbar=1$,
and our convention for the metric signature is timelike $\eta^{\mu\nu}={\rm diag}(+,-,-,-)$.

The outline of this paper is as follows.
In Sec.~\ref{model-basics}, 
our Lorentz-violating model Lagrangian is introduced 
and some of its properties are reviewed. 
Section~\ref{2pointfunction} 
contains a discussion
of
the fermion two-point function within this model,
the correct way to extract the one-particle pole in the Lorentz-violating case,
and the derivation of a general formula for
the corresponding spinor wave-function renormalization factor.
In Sec.~\ref{one-loop}, 
the one-loop radiative corrections to the fermion propagator are evaluated,
the one-particle pole is extracted, 
and the dispersion relation as well as the spinor wave-function renormalization factor are obtained.
Section~\ref{external-LSZ} extends the LSZ formalism 
to the Lorentz-violating case 
and establishes the associated Feynman expansion of the scattering matrix.
In Sec.~\ref{Coulomb-scattering}, 
the formalism developed in this paper
is applied to the example of Coulomb scattering. 
Our summary and an outlook are contained in Sec.~\ref{conclusions}.
Supplemental material is collected in various Appendixes.


\section{Model Basics and Scope}
\label{model-basics}

Our model is based on the 
bare 
gauge-invariant
flat-spacetime 
Lagrange density for single-flavor quantum electrodynamics (QED) 
within the minimal SME: 
\begin{eqnarray}\label{bare-lagrangian-mSME}
\mathcal{L}_{\rm SME} \!& = &\! \half{\it i}\,\overline{\psi}_B\,
{\Gamma}^{\mu}_B
\!\!\stackrel{\;\leftrightarrow}
{D}\!\!{}_{\mu}^B \, {\psi}_B
-\overline{\psi}_B\, M_B \,{\psi}_B-\quar (F_B)^{2}\nonumber\\
&&\! {}-\quar(k^{B}_{F})_{\mu\nu\rho\sigma}F^{\mu\nu}_B F^{\rho\sigma}_B+(k^{B}_{AF})^{\mu}A^{\nu}_B \tilde{F}^B_{\mu\nu}\;.
\end{eqnarray}
The label $B$ denotes bare quantities,
$\psi_B$ is a Dirac spinor
and $F^B_{\mu\nu}=\partial_\mu A^B_\nu-\partial_\nu A^B_\mu$ 
a gauge-field strength.
We have also implemented the conventional notation for 
the U(1)-covariant derivative 
$D^B_{\mu}=\partial_\mu+ie_B A^B_\mu$
and for the dual field-strength tensor 
$\tilde{F}_B^{\mu\nu}=\half \varepsilon^{\mu\nu\rho\sigma} F^B_{\rho\sigma}$.
The Lorentz-violating effects are contained in 
the quantities $(k^{B}_{F})_{\mu\nu\rho\sigma}$ and $(k^{B}_{AF})^{\mu}$
as well as in
the generalized gamma matrices $\Gamma^\mu_B$ and
the generalized mass matrix $M_B$.
The latter are given by the explicit expressions
\begin{eqnarray}\label{covar-Gamma-M-B}
\Gamma_B^\mu\!&=&\!\gamma^\mu+c_B^{\mu\nu}\gamma_\nu+d_B^{\mu\nu}\gamma_5\gamma_\nu+if_B^\mu
+\half g_B^{\lambda\nu\mu}\sigma_{\lambda\nu}+e_B^\mu\,,\nonumber\\
M_B\!&=&\!m_B+a_B^\mu\gamma_\mu+b_B^\mu\gamma_5\gamma_\mu+\half H_B^{\mu\nu}\sigma_{\mu\nu}\,.
\end{eqnarray}
The nondynamical spacetime constant quantities
$(k^{B}_{F})_{\mu\nu\rho\sigma}$,
$(k^{B}_{AF})^{\mu}$,
$a^\mu_B$, 
$b^\mu_B$, 
$c^{\mu\nu}_B$, 
$d^{\mu\nu}_B$, 
$e^\mu_B$, 
$f^\mu_B$, 
$g^{\lambda\mu\nu}_B$, and 
$H^{\mu\nu}_B$
control the type and extent of Lorentz and CPT breakdown.
It has been shown that
this flat-spacetime Lagrangian is multiplicatively renormalizable at one-loop order~\cite{Kostelecky:2001jc}
and that this renormalizability property is maintained in curved spacetimes~\cite{Shapiro}.

The complete one-loop structure of Lagrangian~(\ref{bare-lagrangian-mSME})
would be of interest, 
but lies beyond the scope of this work.
Our present goal is rather to initiate the study of finite radiative corrections
in the presence of Lorentz violation
by highlighting several theoretical issues that
can arise within this context.
For such illustrative purposes,
it seems appropriate to simplify the model~(\ref{bare-lagrangian-mSME})
such that tractability, 
phenomenological importance, 
and theoretical relevance are optimized.
Considerations along these lines are presented next.

A key simplification is setting to zero all Lorentz-violating coefficients, 
with the exception of $c_B^{\mu\nu}$ and $(k_F^B)_{\mu\nu\rho\sigma}$.
We may also take the $c^B_{\mu\nu}$ coefficient to be symmetric
because its antisymmetric piece can be removed from the Lagrangian 
by a field redefinition at leading order~\cite{sme}.
Moreover, 
we will choose $k_F^B$ to be of the form
\begin{equation}\label{special-k_F}
(k_F^B)^{\mu\nu\rho\sigma}=
\half (\eta^{\mu\rho}\tilde{k}_B^{\nu\sigma}-\eta^{\nu\rho}\tilde{k}_B^{\mu\sigma}
-\eta^{\mu\sigma}\tilde{k}_B^{\nu\rho}+\eta^{\nu\sigma}\tilde{k}_B^{\mu\rho})\,,
\end{equation}
where $\tilde{k}_B^{\mu\nu}$ is taken as symmetric, traceless, and given by
\begin{equation}\label{ktilde}
\tilde{k}_B^{\mu\nu}=(k_F^B)^{\mu\alpha\nu}{}_{\alpha}
\end{equation}
We remark that the above choice of Lorentz-violating couplings 
is compatible with the structure of the renormalization constants, 
as will become apparent below.
In particular, 
no additional operators are needed to absorb ultraviolet divergences 
in the perturbative quantum-field expansion of the model.
The above
choice of SME coefficients 
also requires a number of additional considerations,
which we present next.

First, 
we note that Eqs.~(\ref{special-k_F}) and (\ref{ktilde})
are incompatible in spacetime dimensions $d\neq 4$.
When dimensional regularization is employed,
it might then appear that 
this could lead to interpretational difficulties 
with previously determined 
$(k^B_F)_{\mu\nu\rho\sigma}$ and $c^B_{\mu\nu}$ 
counterterms~\cite{Kostelecky:2001jc} 
in the context of minimal subtraction,
affect finite radiative corrections,
or may even be associated with trace anomalies.
However, 
it turns out that 
such spurious issues can be avoided altogether by 
considering a model with $\tilde{k}_B^{\mu\nu}$ in its own right 
rather than as the limit of the full $(k^B_F)_{\mu\nu\rho\sigma}$ Lagrangian.
Throughout this work, 
we follow this latter, independent interpretation of our model.

Second,
the identification of observables
in Lorentz-violating field theories 
requires special care
due various types field redefinitions and reinterpretions~\cite{fieldredef}.
In the present case, 
it turns out that
the $c^{\mu\nu}$ and $\tilde{k}^{\mu\nu}$ coefficients
are observationally indistinguishable at leading order
in any fermion--photon system;
only their difference $2c^{\mu\nu}-\tilde{k}^{\mu\nu}$ can be measured 
within the context of Lagrangian~\rf{bare-lagrangian-mSME}.
This feature has previously been discussed  
from various perspectives~\cite{KM02,rescaling,hohensee}. 
For example, 
suitable coordinate rescalings
can eliminate the $c^{\mu\nu}$ coefficient
in favor of $\tilde{k}^{\mu\nu}$,
or vice versa.
Such coordinate redefinitions
can be exploited to simplify calculations.
In what follows, 
however,
we will for the most part avoid the choice of a particular coordinate scaling 
by keeping both $c^{\mu\nu}$ and $\tilde{k}^{\mu\nu}$ nonzero. 
This will provide an independent partial test of our results,
since coordinate-scalar expressions for physically observable radiative effects 
should only depend on $2c^{\mu\nu}-\tilde{k}^{\mu\nu}$.

Third, 
we also note that 
various experimental investigations have sought to  constrain
$2c^{\mu\nu}-\tilde{k}^{\mu\nu}$.
In particular,
measurements have been performed in the context of
resonance cavities~\cite{cavities},
kinematical threshold studies at colliders~\cite{hohensee},
synchrotron radiation~\cite{synchrotron},
Compton-edge investigations in electron--photon scattering~\cite{graal},
and astrophysical observations~\cite{astro}. 
Through these investigations, 
all components of $2c^{\mu\nu}-\tilde{k}^{\mu\nu}$ 
are currently obeying bounds at the levels of $10^{-13}\ldots10^{-17}$. 
At present, 
$2c^{\mu\nu}-\tilde{k}^{\mu\nu}$ 
nevertheless remains the parameter combination in
Lagrangian~\rf{bare-lagrangian-mSME} with the weakest experimental limits 
providing additional phenomenological justification for
dropping the other SME coefficients from our analysis.
We finally mention that 
further constraints on $2c^{\mu\nu}-\tilde{k}^{\mu\nu}$ may, 
for example, 
also be determined with spectroscopic studies of hydrogen~\cite{adkins}
and ultrahigh-energy photon-shower measurements~\citep{Rubtsov:2012kb}.

Paralleling the conventional case, 
perturbative calculations within the present model 
are conveniently performed by fixing a gauge and 
allowing for the need to regularize infrared divergences.
For the general description of massive Lorentz-violating photons,
a modified Stueckelberg procedure has recently been developed~\cite{Stueckelberg}.
It essentially consists of amending any Lorentz-violating QED Lagrange density by
\begin{equation}
\Delta{\cal L}=-\half\xi^{-1}(\partial_\mu \tilde\eta_B^{\mu\nu}\!A^B_\nu)^2
+\half m_\gamma^2\,A^B_\mu \tilde\eta_B^{\mu\nu}\!A^B_\nu\,,
\label{deltaL}
\end{equation}
where $\xi$ denotes a gauge parameter and 
$m_\gamma$ parametrizes the gap in the photon dispersion relation.
The tensorial structure $\tilde\eta_B^{\mu\nu}\equiv\eta^{\mu\nu}+\delta\tilde\eta_B^{\mu\nu}$ 
can involve small,
but otherwise arbitrary Lorentz-breaking contributions $\delta\tilde{\eta}^{\mu\nu}$. 
Note that both the $\xi$ and the $m_\gamma$ term 
need to contain the {\em same} tensor $\tilde\eta_B^{\mu\nu}$~\cite{Stueckelberg}.

The choice of $\delta\tilde\eta_B^{\mu\nu}$ in the present context 
needs to be compatible with the specific purpose
of the $m_\gamma$ term as a regulator: 
no Lorentz violation in addition to $c$ and $\tilde{k}$ should be introduced.
One obvious possibility would be
the Lorentz-symmetric choice $\delta\tilde\eta_B^{\mu\nu}=0$. 
Another possibility is to match the Lorentz-violating structure
of the effective kinetic term of the photon.
At leading order in $c$ and $\tilde{k}$, 
this kinetic term depends on the combination $\eta^{\mu\nu}+\tilde{k}_B^{\mu\nu}$.
If radiative corrections are included, 
the $c_B^{\mu\nu}$ coefficient may also appear, 
but only in the combination $(2c_B^{\mu\nu}-\tilde{k}_B^{\mu\nu})$, 
as discussed above.
These considerations 
suggest the following choice for $\delta\tilde{\eta}^{\mu\nu}$:
\begin{equation}
\delta\tilde\eta_B^{\mu\nu}=
\tilde{k}_B^{\mu\nu}+f(2c_B^{\mu\nu}-\tilde{k}_B^{\mu\nu})\,.
\label{delta_eta}
\end{equation}
Here, 
$f$ is a free multiplicative coefficient that 
may for example be chosen to match the radiative corrections 
to free-photon propagation.
In the present work,
it will be convenient to select the choice~\rf{delta_eta}.
Since our primary focus is the fermion two-point function,
we will set $f=0$.

Altogether, the above considerations lead to the following
bare Lagrange density~\cite{ghostfn}:
\begin{align}
\mathcal{L} &= 
  \bar \psi_B \left[i \left(\gamma^\mu + c^{\mu\nu}_B\gamma_\nu\right) 
  \left(\prt_\mu + i e_B A^B_\mu \right) - m_B \right] \psi_B  \nonumber\\
 &\quad{} - \frac14 \tilde\eta_B^{\mu\nu} \tilde\eta_B^{\alpha\beta}
   F^B_{\mu\alpha} F^B_{\nu\beta} 
   - \frac{1}{2\xi} \left(\prt_\mu \tilde\eta_B^{\mu\nu} A^B_\nu\right)^2  \nonumber\\
 &\quad{} + \frac{m_\gamma^2}2 \, A^B_\mu \tilde\eta_B^{\mu\nu} A^B_\nu\,.
\label{bare-model-lagrangian}
\end{align}

The next step is to define finite fields and couplings. 
To this end, 
we employ the usual multiplicative renormalization procedure 
with its Lorentz-violating generalization established in Ref.~\cite{Kostelecky:2001jc}:
\begin{equation}
\renewcommand{\arraystretch}{1.5}
\begin{array}{c}
\psi_B=\sqrt{Z_\psi}\,\psi\,, \qquad A_B^\mu=\sqrt{Z_A}\,A^\mu\,,\\
m_B=Z_m\, m\,, \qquad e_B=Z_e\, e\,,\\
c_B^{\mu\nu}=(Z_c)^{\mu\nu}{}_{\alpha\beta}\,c^{\alpha\beta}
\equiv (Z_c c)^{\mu\nu}\,, \\
\tilde{k}_B^{\mu\nu}=(Z_k)^{\mu\nu}{}_{\alpha\beta}\,\tilde{k}^{\alpha\beta}
\equiv (Z_k \tilde k)^{\mu\nu}\,.
\end{array}
\label{physical-bare}
\end{equation} 
Adopting Feynman gauge ($\xi=1$),
working in $4-\epsilon$ dimensions,
and employing minimal subtraction, 
we have at one-loop order \cite{Kostelecky:2001jc}:
\begin{eqnarray}\label{renormalization-constants1}
& Z_m=1-{\displaystyle{3e^2\over8\pi^2\epsilon}}\,, 
\quad Z_e = 1+{\displaystyle{e^2\over12\pi^2\epsilon}}\,,& \\
& Z_\psi=1-{\displaystyle{e^2\over8\pi^2\epsilon}}\,,
\quad Z_A=1-{\displaystyle{e^2\over6\pi^2\epsilon}}\,,&
\label{renormalization-constants2}\\
&(Z_c c)^{\mu\nu}=c^{\mu\nu}+{\displaystyle{e^2\over6\pi^2\epsilon}}
\left(2c^{\mu\nu}-\tilde{k}^{\mu\nu}\right)\,,&
\label{renormalization-constants3}\\
&(Z_k \tilde k)^{\mu\nu}=\tilde{k}^{\mu\nu}+{\displaystyle{e^2\over6\pi^2\epsilon}}
\left(\tilde{k}^{\mu\nu}-2c^{\mu\nu}\right)\,.&
\label{renormalization-constants4}
\end{eqnarray}
We remark that these expressions 
are compatible with 
the usual QED Ward--Takahashi identity $Z_e\sqrt{Z_A}=1$. 
In terms of the above physical couplings and fields,
our model Lagrange density reads
\begin{align}
\mathcal{L}_{c\tilde{k}} &=
Z_\psi \bar \psi \Bigl[ i \Bigl( \gamma^\mu + 
	(Z_c c)^{\mu \nu} \gamma_\nu\Bigr)\nonumber\\ 
&\qquad\qquad{}\times\left(\prt_\mu + iZ_e\sqrt{Z_A} eA_\mu\right)-Z_m m\Bigr]\psi \nonumber\\
&\quad{}-\frac{Z_A}{4} \left( \eta^{\mu \nu} + (Z_k \tilde k)^{\mu \nu}\right)
	\left(\eta^{\alpha\beta}+(Z_k \tilde k)^{\alpha\beta}\right) 
	F_{\mu \alpha}F_{\nu \beta}\nonumber\\
&\quad{}-\frac{Z_A}{2\xi}\left(\bigl(\eta^{\mu\nu}+(Z_k \tilde k)^{\mu\nu}\bigr)
	\prt_\mu A_\nu \right)^2\nonumber\\
&\quad{}+ \frac{Z_A m_\gamma^2}2 A_\mu\left(\eta^{\mu\nu}+
	 (Z_k \tilde k)^{\mu\nu}\right)A_\nu \,.
	 \label{model-lagrangian-renorm}
\end{align}


\section{The fermion two-point function}
\label{2pointfunction}

Our main objective being the treatment of external fermion
states in the context of Lorentz-violating field theory,
we will turn in this section to the general procedure of 
determining the on-shell limit of the two-point function 
in our model~\rf{model-lagrangian-renorm}.
We are interested in particular in the Lorentz-breaking
radiative corrections to this limit.
In Lorentz-symmetric field theory, 
one extracts from the general off-shell two-point function 
the one-particle pole and its residue, 
which respectively determine 
the asymptotic single-particle solutions 
and the wave-function renormalization coefficient. 
Lorentz invariance strongly restricts the form of 
the different types of terms that 
can occur in the fermion two-point function. 
In the presence of the Lorentz-violating parameters 
$c^{\mu\nu}$ and $\tilde{k}^{\mu\nu}$,
this procedure has to be generalized 
because more general terms can (and do) occur, 
the only fundamental restriction being that 
they are observer Lorentz scalars. 
What makes this generalization particularly nontrivial 
is the fact that 
some of the terms involving gamma matrices 
become noncommuting, 
an effect that 
does not occur in the usual Lorentz-symmetric case.

In a recent study~\cite{Potting:2011yj}, 
a generalization of the K\"all\'en--Lehmann
spectral representation was derived for scalar and fermion field theories
in the presence of a Lorentz-violating background of the type considered
in this work.
We will use the form derived for the one-particle pole in that work
as a guiding principle for extracting the one-particle fermion pole in the present analysis.

\begin{center}
\begin{figure}[h]
\includegraphics[width=0.85\hsize]{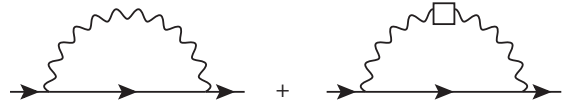}
\caption{$\mathcal{O}(\alpha, \tilde k)$ loop contributions to 
the fermion self-energy in the first perturbation scheme 
with $c^{\mu\nu} = 0$.
The single solid and wavy lines denote conventional Lorentz-symmetric electron and photon propagators, respectively. The box represents the Lorentz-violating $\tilde{k}$ insertion~\cite{footnote-1}.}
 \label{pertscheme1}
\vskip-20pt
 \end{figure}
\end{center}

We begin our general discussion of the two-point function 
with a few general remarks about the choice of perturbation scheme.
(An actual perturbative calculation of this function 
will have to wait until the next section.)
To set up perturbation theory 
for calculating the radiative corrections to the two-point function 
we have to make 
a suitable choice of a zeroth-order system with known solutions,
such that the remaining piece can be considered a small perturbation 
relative to this zeroth-order system. 
While usually one takes as the zeroth-order system the full quadratic
part of the action,
in the case at hand 
there are at least two reasonable choices one might consider.

In the first scheme, one defines as
a basis the renormalized quadratic Lagrangian of
the conventional Lorentz-symmetric case.
All Lorentz-violating contributions to the
Lagrangian are then taken as perturbations.
One can, for example, 
use the minimal-subtraction scheme
to define the counterterms 
(either Lorentz-symmetric or Lorentz-violating).

In the second scheme, one defines as a basis
the full renormalized quadratic Lagrangian,
including the Lorentz-violating part.
The perturbations are then just the nonquadratic contributions
to the Lagrangian.
For the latter, it is convenient to join the corresponding
Lorentz-symmetric and Lorentz-violating vertices in the same diagram.
One can do this also for the counterterms.

A key difference between the two schemes 
concerns their kinematical features, 
which is best explained by an example.
Consider the one-loop fermion self energy.
In the conventional Lorentz-symmetric case, 
energy--momentum kinematics prohibits
the internal photon and fermion lines 
from going on-shell simultaneously
for physical incoming momenta.
Let us next look at the leading-order Lorentz-violating generalization of this process
in the above two schemes
with focus on the special case with a photon $\tilde{k}$ coefficient only.

In the first scheme, 
this process involves the conventional diagram 
plus another version of the diagram with a $\tilde{k}$ insertion
on the photon line 
represented by a box in Fig.~\ref{pertscheme1}.
Both of these diagrams exhibit 
the same Lorentz-symmetric propagators 
and the same Lorentz-symmetric dispersion relation 
for external momenta.  
The energy--momentum kinematics 
is therefore unchanged relative to the conventional case.
In particular,
the internal photon and fermion propagators cannot go on-shell simultaneously
for physical incoming momenta
at this order in perturbation theory.

\begin{center}
\begin{figure}[h]
\includegraphics[width=0.40\hsize]{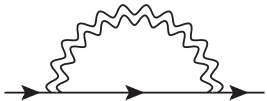}
\caption{$\mathcal{O}(\alpha, \tilde k)$ loop contributions to the fermion self energy in the second perturbation scheme with $c^{\mu\nu} = 0$.
The single solid line denotes the conventional Lorentz-symmetric electron propagator. The double wavy line represents the full Lorentz-violating photon propagator including all tree-level $\tilde{k}$ effects~\cite{footnote-1}.}
\label{pertscheme2}
\vskip-20pt
\end{figure}
\end{center}

In the second scheme, 
there is a single diagram analogous to the conventional one,
but with the usual photon propagator replaced by
a Lorentz-violating propagator containing $\tilde{k}$, represented
by a double photon line in Fig.~\ref{pertscheme2}.
At this order, 
the dispersion relation of the incoming momentum 
remains Lorentz symmetric,
but the $\tilde{k}$ propagator modifies the internal kinematics 
of the self-energy diagram. 
In particular, 
the internal photon and fermion lines can now go on-shell simultaneously
for physical incoming momenta
in certain regimes.
This is evident from the well-established result~\cite{AltCer} that
photons that are slowed down by certain values of $\tilde{k}$ 
lead to vacuum Cherenkov radiation
at ultrahigh energies:
a real free fermion can now emit a real photon.
The nonzero cross section for this effect
is then directly related to imaginary contributions
to the fermion self-energy 
via the optical theorem.
Note that 
imaginary contributions are absent in the first scheme.

Such Cherenkov instabilities are relatively rare:
they do not occur for all Lorentz-violating coefficients
and are in any case only present at ultrahigh energies in our model.
They correspond exactly to the two-particle
regime discussed in Ref.~\cite{Potting:2011yj},
where the K\"all\'en--Lehmann representation of the fermion
two-point function was analyzed in the presence of a $c^{\mu\nu}$-type
Lorentz-violating perturbation.
As was shown there, for ultrahigh momenta the two-point function
can pass from a stable one-particle regime into an unstable two-particle
regime and provoke a Cherenkov-type decay.
In the case at hand, 
the latter consists of the fermion plus a photon.

In this work,
we are primarily interested in the properties of external states 
{\em before} such rare instabilities lead to their decay. 
We therefore omit imaginary contributions to the fermion self-energy.
If needed, 
the physics of such instabilities may still be included subsequently,
for example via cut diagrams using Cutkosky's rules~\cite{Cut}.
For the above purpose, 
which disregards instabilities, 
the two schemes become equivalent---a fact we have verified explicitly
at one-loop order.
For definiteness, 
we present our analysis in the second scheme.
It has the advantage of being more economical, 
as the number of diagrams is considerably reduced. 
In particular, 
the diagrams in this scheme 
are in one-to-one correspondence 
with the diagrams in the Lorentz-symmetric case. 
Moreover, 
in the present context 
this scheme is free of momentum-routing ambiguities.
Although the prescription for calculating the corresponding amplitude
is more involved  due to the Lorentz-violating coefficients, 
we have found that 
the full calculation is easier to carry out in the second scheme 
and is also less prone to error.

Explicitly, the second scheme implies that our model
Lagrange density~\rf{model-lagrangian-renorm} is split
into the following three pieces:
\begin{equation}
\mathcal{L}_{c\tilde{k}}=\mathcal{L}_0+\mathcal{L}_1+\mathcal{L}_2\,,
\label{model-lagrangian-split}
\end{equation}
where
\begin{align}
\mathcal{L}_0 &=
\bar \psi \left[ i \left( \gamma^\mu + c^{\mu \nu} \gamma_\nu \right) \prt_\mu - m \right] \psi  \nonumber\\
&\>{}-\frac{1}{4}\left( \eta^{\mu \nu} + \tilde k^{\mu \nu} \right) 
	\left( \eta^{\alpha \beta} + \tilde k^{\alpha \beta} \right) F_{\mu \alpha}F_{\nu \beta}  \nonumber\\
&\>{}-\frac{1}{2 \xi} \left(\prt_\mu A^\mu + \tilde k^{\mu \nu} \prt_\mu A_\nu \right)^2
+\frac{m_\gamma^2}{2}A_\mu\left(\eta^{\mu\nu}+\tilde k^{\mu\nu}\right)A_\nu
\,,\label{L0}\\
\mathcal{L}_1 &= \bar \psi \left[ -e \left( \gamma^\mu + c^{\mu \nu} \gamma_\nu \right) A_\mu \right] \psi\,,\label{L1}
\end{align}
and
\begin{align}
\mathcal{L}_2 &= \bar\psi\Bigl[\bigl((Z_\psi - 1)\eta^{\mu\nu} + Z_\psi(Z_c\,c)^{\mu\nu} - c^{\mu \nu}\bigr)
		i \gamma_\nu \left(\prt_\mu + i e A_\mu \right)\nonumber\\
&\qquad\qquad{}- ( Z_\psi Z_m - 1) m \Bigr] \psi \nonumber\\
&\quad{}- \frac14\biggl[\left(Z_A -1\right) \eta^{\mu \nu} \eta^{\alpha \beta} +  
		2 \left( Z_A ( Z_k\, \tilde k )^{\mu \nu} - \tilde k ^{\mu \nu}\right)
		\eta^{\alpha \beta} \nonumber\\
&\qquad\qquad {}+  \Bigl( Z_A ( Z_k \,\tilde k)^{\mu\nu} 
		(Z_k\,\tilde k)^{\alpha\beta} - \tilde k^{\mu\nu} \tilde k^{\alpha\beta} \Bigr) \biggr]\nonumber\\
&\qquad\qquad{}\times\left( F_{\mu \alpha}F_{\nu \beta} + \frac{2}{\xi}
\left(\prt_\mu A_\nu \right)\left(\prt_\alpha A_\beta \right) \right)\nonumber\\
&\quad{}+\frac{m_\gamma^2}2 A_\mu \left[(Z_A-1)\eta^{\mu\nu}
       + Z_A(Z_k\,\tilde k)^{\mu\nu}-\tilde k^{\mu\nu}\right] A_\nu
       \,.\label{L2}
\end{align}

The corresponding Feynman rules,
which are collected in Appendix~\ref{2pointrules},
now facilitate an order-by-order calculation 
of our model's two-point function
\begin{equation}\label{proper-two-point-function}
\Gamma^{(2)}(p)=\Gamma^\mu p_\mu-m-\Sigma(p^\mu)\,.
\end{equation}
Paralleling the conventional perturbative determination of this function,
$\Sigma$ denotes the contribution of the
one-particle irreducible Feynman diagrams. 
The summation procedure for these diagrams,
which does not rely on Lorentz symmetry,
closely parallels the conventional case 
and leads directly to Eq.~(\ref{proper-two-point-function}).

Before proceeding with an actual one-loop calculation, 
it is instructive to determine the general structure of $\Sigma$.
This structure is constrained by the requirement of 
coordinate independence,
which dictates that $\Sigma$ 
can only depend on coordinate scalars 
formed by contractions of 
model parameters, external momenta, and gamma matrices.
In particular, 
one can consider the propagator as an effective function 
of these contracted Lorentz scalars.
This represents a direct generalization of the conventional case,
where a single independent scalar, $\slashed p$, 
can be formed 
and the propagator can be treated as depending on $\slashed p$.

With these considerations in mind, 
we may decompose $\Sigma$ as
\begin{equation}\label{SigmaDecomp}
\Sigma=
\Sigma_{\rm LI}(\slashed p)
+\Sigma_{\rm LV}(p^2, c^{p}_\gamma, \,\tilde{k}^{p}_\gamma)
+\delta\Sigma
(p^\mu,c^{\mu\nu},\tilde{k}^{\mu\nu})\,,
\end{equation}
where we have defined 
$c^{p}_\gamma\equiv c^{\mu\nu}\gamma_\mu p_\nu$ and
$\tilde{k}^{p}_\gamma\equiv \tilde{k}^{\mu\nu}\gamma_\mu p_\nu$.
In this decomposition,
$\Sigma_{\rm LI}(\slashed p)$ denotes the Lorentz-symmetric contributions equivalent to the conventional diagrams;
it can thus only be a function of $\slashed p$, 
as usual:
\begin{equation}\label{SigmaLI}
\Sigma_{\rm LI}(\slashed p)=f_0(p^2)m+f_1(p^2)\,\slashed p\,, 
\end{equation} 
where both $f_0(p^2)$ and $f_1(p^2)$ 
are understood to depend on the 
fine-structure constant $\alpha=e^2/4\pi$ and 
the square of the 4-momentum $p^2$.
The remaining two terms 
involve deviations from Lorentz symmetry,
so we will describe them in more detail.

The second term $\Sigma_{\rm LV}(p^2, c^{p}_\gamma, \tilde{k}^{p}_\gamma)$ 
contains all those Lorentz-violating terms 
with a gamma-matrix structure that
is already present in the fermion Lagrange density 
(i.e., a Lorentz-breaking symmetric traceless 2-tensor 
contracted with a gamma matrix and a single momentum factor).
For example, 
$\Sigma_{\rm LV}$ includes the counterterms for $c^{\mu\nu}$ and $\tilde{k}^{\mu\nu}$ 
together with the corresponding (regulated) infinities they cancel 
to yield an ultraviolet finite expression:
\begin{equation}\label{SigmaLV}
\Sigma_{\rm LV}(p^2, c^{p}_\gamma, \tilde{k}^{p}_\gamma)=
f_2^c(p^2)\, c^{p}_\gamma +f_2^{\tilde k}(p^2)\,\tilde{k}^{p}_\gamma\,,
\end{equation}
where $f_2^c(p^2)$ and $f_2^{\tilde k}(p^2)$ 
depend only on $\alpha$ and $p^2$ 
since we are working at leading order in Lorentz violation.
Explicit expressions for $f_2^c(p^2)$ and $f_2^{\tilde k}(p^2)$ 
can in principle be determined within
perturbation theory to any given order in $\alpha$.
Below we will calculate $f_2^c(p^2)$ and $f_2^{\tilde k}(p^2)$ at one loop, 
i.e., at ${\cal O}(\alpha)$. 
Initially, 
$f_2^c(p^2)$ and $f_2^{\tilde k}(p^2)$
may also depend on an infrared regulator and 
an arbitrary mass scale introduced
by the chosen ultraviolet regularization procedure.
But 
a consistent treatment of infrared effects
and the renormalization conditions 
should remove free parameters 
from $\Sigma_{\rm LV}(p^2, c^{p}_\gamma, \tilde{k}^{p}_\gamma)$.
 
The remaining term $\delta\Sigma(p^\mu,c^{\mu\nu},\tilde{k}^{\mu\nu})$
contains novel Lorentz-breaking structures that
are {\em not} already present in the original Lagrange 
density~\rf{model-lagrangian-renorm}.
Like the second term, 
it must involve combinations of 
$p^\mu$ factors, 
$\gamma$ matrices, 
and---since we are working at linear order in Lorentz violation---a 
single $c^{\mu\nu}$ or $\tilde{k}^{\mu\nu}$ coefficient.
Up to factors consisting of powers of $p^2$, 
a multitude of terms can be constructed 
that satisfy these requirements. 
They are
\begin{eqnarray}\label{allterms}
&&\hspace{-7mm}c^{\mu\nu}p_\mu p_\nu\,,\quad \slashed p\, c^{\mu\nu}p_\mu p_\nu\,,
\quad\gamma^5 c^{\mu\nu}p_\mu p_\nu\,,\quad
\gamma^5\slashed p\, c^{\mu\nu}p_\mu p_\nu
\,,\nonumber\\
&&\hspace{-7mm}\gamma^5 c^{\mu\nu}\gamma_\mu p_\nu\,,
\quad\sigma^{\lambda\mu}c_\lambda{}^\nu p_\mu p_\nu\,,
\quad\gamma^5\sigma^{\lambda\mu}c_\lambda{}^\nu p_\mu p_\nu\,,
\end{eqnarray}
as well as an additional seven terms with $c^{\mu\nu}$ replaced by $\tilde{k}^{\mu\nu}$
\cite{footnote0}.

The above list~(\ref{allterms}) can be constrained further by noting that 
electromagnetic interactions preserve C, P, and T.
Quantum corrections linear in $c^{\mu\nu}$ and $\tilde{k}^{\mu\nu}$
must exhibit the same discrete symmetries
as the original Lorentz-violating operators.
This fact 
together with our scope set out earlier
(i.e., omitting instabilities and thus non-Hermitian expressions) 
leaves only the first two terms in the list~\rf{allterms} 
and their $\tilde{k}^{\mu\nu}$ analogues~\cite{footnote1}.
For this reason, 
$\delta\Sigma(p^\mu,c^{\mu\nu},\tilde{k}^{\mu\nu})$
can only depend on
$c^{p}_p\equiv c^{\mu\nu}p_\mu p_\nu$ and
$\tilde{k}^{p}_p\equiv \tilde{k}^{\mu\nu}p_\mu p_\nu$, 
and we may write
\begin{align}
\label{deltaSigma}
\delta\Sigma(p^\mu,c^{p}_p,\tilde{k}^{p}_p)&=
f_3^c(p^2)\, \frac{c^{p}_p}m
+f_4^c(p^2)\, \frac{\slashed pc^{p}_p}{m^2} \nonumber\\
&\qquad{}+f_3^{\tilde k}(p^2)\,\frac{\tilde{k}^{p}_p}m
+f_4^{\tilde k}(p^2)\,\frac{\slashed p\tilde{k}^{p}_p}{m^2}\,.
\end{align}
Here, 
we have introduced the dimensionless functions 
$f_3^c(p^2)$,
$f_4^c(p^2)$,
$f_3^{\tilde k}(p^2)$, and
$f_4^{\tilde k}(p^2)$,
which can in principle be calculated within perturbation theory
to any given order in $\alpha$.
These functions may initially still contain infrared regulators,
which can presumably be removed by a soft-photon treatment.
Disregarding the aforementioned possibility of high-energy non-Hermitian contributions,
Eqs.~\rf{SigmaDecomp}, \rf{SigmaLI}, \rf{SigmaLV}, and \rf{deltaSigma}
determine the full off-shell structure of the fermion two-point function 
in our $c\tilde{k}$ model~\rf{model-lagrangian-renorm}
at all orders in $\alpha$ and 
at linear order in Lorentz violation. 

Before deriving explicit expressions for 
the scalar functions appearing 
in the corrections~\rf{SigmaLI}, \rf{SigmaLV}, and~\rf{deltaSigma}---a
task to which we will turn in Sec.~\ref{one-loop}---it
is instructive to construct a general procedure for 
extracting the on-shell external-leg physics 
determined by the structure of these corrections.
In general, 
external tree-level momentum-space Dirac spinors $w_0$ 
satisfy $P_0 w_0=0$, 
where $P_0$ denotes the free Dirac operator of the theory 
(e.g., $P_0=\slashed p -m$ in the conventional case). 
The external states $w$ in the fully interacting theory 
must then satisfy an equation of the structure 
$(P_0 + \delta P' )w=0$, 
where $\delta P'$ is a small correction to $P_0$~\cite{PertCom}.
It is customary to rearrange this equation
such that all terms proportional the matrix $\slashed p$
are removed from $\delta P'$~\cite{ZCom}. 
This yields $\mathcal Z_R^{-1}(P_0 + \delta P)w=0$, 
where $\mathcal Z_R$ is some scalar function
and $\delta P$ is properly adjusted relative to $\delta P'$.
For sufficiently small interactions, 
$\mathcal Z_R=1+\delta \mathcal Z_R$
should remain close to unity 
and is thus regular for on-shell momenta.

The expression $\bar P\equiv P_0 + \delta P$ 
can be interpreted as 
the effective single-particle Dirac operator in the fully interacting theory, 
which governs the propagation of 
external states.
Standard arguments
now directly imply that 
the momentum-space Green function associated with $\bar P$
{\em must} have the structure $\mathcal Z_R \bar P{}^{-1}$.
In addition to this one-particle pole,
the two-point function may also contain 
additional off-shell effects and multiparticle physics, 
which can be included via a general term $R$ 
that remains regular in the vicinity of the pole:
\begin{equation}\label{GenPole}
\Gamma^{(2)}(p)^{-1}=
\mathcal Z_R \bar P{}^{-1}-R\,.
\end{equation}
This result is fully consistent with both the original K\"allen--Lehmann representation~\cite{KallenLehmann}
and its recent generalization to Lorentz-violating theories~\cite{Potting:2011yj}.

The reasoning leading Eq.~(\ref{GenPole}) 
leaves undetermined the detailed structures of
$\mathcal Z_R$, $\delta P$, and $R$.
However, 
perturbative expressions for these quantities 
can be determined.
For example, 
one may compare 
an explicit loop-expansion result for $\Gamma^{(2)}(p)$
to the following form of Eq.~(\ref{GenPole}),
\begin{equation}\label{InvGenPole}
\Gamma^{(2)}(p)=\mathcal Z_R^{-1} \bar P
+\bar P\big[ \mathcal Z_R^{-2}R(\openone-\mathcal Z_R^{-1} \bar PR)^{-1}\big]\bar P\,.
\end{equation}
Note that up to this point,
the above procedure, 
and in particular the result~(\ref{InvGenPole}),
are general 
and do not rely on Lorentz invariance.
Symmetry considerations typically enter in the next step,
when a general ansatz for $\delta P$ is posited
and the free parameters contained in this ansatz
are determined via 
comparison to the loop expression for $\Gamma^{(2)}(p)$. 

As an example, 
let us briefly review the essence of the conventional QED case.
A perturbative evaluation of the two-point function yields
\begin{equation}\label{LI2point}
\Gamma^{(2)}_{\rm LI}=A(p^2)\slashed p+M(p^2)\openone=
A(\slashed p\slashed p)\slashed p-M(\slashed p\slashed p)\openone\,,
\end{equation}
with explicit expressions for $A(p^2)$ and $M(p^2)$ 
that depend on the order in pertubation theory~\cite{LIcomment}. 
One then considers the ansatz 
\begin{equation}\label{LIansatz}
\bar P=\slashed p -m_{\rm ph}
\end{equation}
for the dispersion relation~\cite{LIcomment}, 
where $m_{\rm ph}$ is a free parameter to be determined.
Note that this is the most general form of $\bar P$
that represents a correction to the tree-level case
and exhibits the canonical normalization of $\slashed p$, 
as discussed above.

With this ansatz,
we may rewrite $\Gamma^{(2)}_{\rm LI}$ in Eq.~(\ref{LI2point}) 
as a function of $\bar P$ rather than $\slashed p$:
\begin{eqnarray}\label{LIansatzP}
\Gamma^{(2)}_{\rm LI}(\bar P) & = &
A(\bar P{}^{2}+2m_{\rm ph}\bar P+m_{\rm ph}^2)\bar P\nonumber\\
&&{}+m_{\rm ph}A(\bar P{}^{2}+2m_{\rm ph}\bar P+m_{\rm ph}^2)\openone\nonumber\\
&&{}-M(\bar P{}^{2}+2m_{\rm ph}\bar P+m_{\rm ph}^2)\openone\,.
\end{eqnarray}
The on-shell condition $\Gamma^{(2)}_{\rm LI}(\bar P=0)=0$ 
yields an implicit relation for $m_{\rm ph}$
\begin{equation}\label{LIdetermination}
0=m_{\rm ph}A(m_{\rm ph}^2)-M(m_{\rm ph}^2)\,,
\end{equation}
which can be employed to determine the physical mass.

Expanding $\Gamma^{(2)}_{\rm LI}(\bar P)$
around the pole $\bar P=0$ gives
\begin{equation}
\Gamma^{(2)}_{\rm LI}(\bar P)=
\Gamma^{(2)}_{\rm LI}{}'(0)\,\bar P
+\bar P\Big[\sum_{n=2}^\infty \frac{\Gamma^{(2)}_{\rm LI}{}^{(n)}(0)}{n!}\,\bar P{}^{n-2}\Big]\bar P\,,
\end{equation}
where the zeroth-order term in the expansion vanishes 
by virtue of Eq.~(\ref{LIdetermination}).
Comparison with the general result~(\ref{InvGenPole}) 
then establishes
\begin{eqnarray}\label{LIpoleexpansion}
\hspace{-2em}\mathcal Z_R^{-1}&=&\Gamma^{(2)}_{\rm LI}{}'(0)\nonumber\\
&=&
2m_{\rm ph}^2\big[
A'(m_{\rm ph}^2)
-M'(m_{\rm ph}^2)\big]
+A(m_{\rm ph}^2)\,.
\end{eqnarray}
It is now apparent that
Eqs.~(\ref{LIdetermination}) and~(\ref{LIpoleexpansion})
completely fix the expression for pole.
Note that in the above Lorentz-symmetric situation,
the only nontrivial Dirac-matrix structure is $\slashed p$, 
so that no matrix-ordering issues arise.

In the present Lorentz-violating case, 
we may follow a similar line of reasoning, 
albeit with generalized versions of the above Eqs.~(\ref{LI2point}) and~(\ref{LIansatz}).
Our previous result 
for the general structure of our model's two-point function,
which is summarized in Eqs.~(\ref{proper-two-point-function})--(\ref{deltaSigma}), 
may be recast into the following form:
\begin{equation}
\Gamma^{(2)}=
A(p^2\!, c^p_p,\tilde{k}^p_p)\slashed p+
C(p^2
)c^p_\gamma
+K(p^2
)\tilde{k}^p_\gamma
-M(p^2\!,c^p_p,\tilde{k}^p_p)\,.
\label{LV-expansion1}
\end{equation}
Using our previous definitions, 
we have at leading order in Lorentz violation:
\begin{eqnarray}\label{ACKM_Def}
\hspace{-8mm}A\!&=&\! 1-f_1(p^2)-f_4^c(p^2)\frac{c^p_p}{m^2}
-f_4^{\tilde k}(p^2)\frac{\tilde{k}^p_p}{m^2}\,, \nonumber\\
\hspace{-8mm}C\!&=&\! 1-f_2^c(p^2)\,, \nonumber\\
\hspace{-8mm}K\!&=&\! -f_2^{\tilde k}(p^2)\,, \nonumber\\
\hspace{-8mm}M\!&=&\! m\big[1+f_0(p^2)\big]+f_3^c(p^2)\frac{c^p_p}m
+f_3^{\tilde k}(p^2)\frac{\tilde{k}^p_p}m\,.
\end{eqnarray}
Note that the presence of the Lorentz-breaking parameters
$c^{\mu\nu}$ and $\tilde{k}^{\mu\nu}$ leads to two new features
relative to the Lorentz-symmetric expression~(\ref{LI2point}).
First, 
the coefficient functions $A$ and $M$ 
can now also depend $c^p_p$ and $\tilde{k}^p_p$;
these are the only coordinate scalars in addition to $p^2$ that
can be formed at leading order in Lorentz violation.
Second, 
two additional gamma-matrix structures, 
namely $c^p_\gamma$ and $\tilde{k}^p_\gamma$,
and their respective coefficient functions 
$C(p^2,c^p_p,\tilde{k}^p_p)$ and $K(p^2,c^p_p,\tilde{k}^p_p)$ 
can now be formed.
As for the Lorentz-invariant case,
the detailed expressions for $A$, $M$, $C$, and $K$ 
depend on the order in $\alpha$ under consideration.

Next, 
we need the generalization of the ansatz~(\ref{LIansatz}).
Employing the results of Ref.~\cite{Potting:2011yj} 
at leading order in $c^{\mu\nu}$ and $\tilde{k}^{\mu\nu}$, 
we find the most general form for the pole to be
\begin{equation}
\bar P=\slashed p-\bar m
+\bar x c^p_\gamma+\bar y\tilde k^p_\gamma\,.
\label{LV-pole}
\end{equation}
Here, 
the coefficient functions $\bar m$, $\bar x$, and $\bar y$ 
do not depend on $p^2$. 
This is intuitively reasonable, 
since on-shell we may replace 
$p^2\to m_{\rm ph}^2+{\cal O}(c^p_p,\tilde{k}^p_p)$.
In any case, 
this follows rigorously from the results in Ref.~\cite{Potting:2011yj}.
This means 
we can take $\bar x$ and $\bar y$ to be free constants 
to be determined later. 
Similarly, 
\begin{equation}
\bar m = m_{\rm ph}+m_c c^p_p+m_k\tilde k^p_p\,, 
\label{m-first-order}
\end{equation}
with $m_{\rm ph}$, $m_c$, and $m_k$ 
momentum-independent parameters 
to be determined below.
Note that as opposed to the usual Lorentz-symmetric ansatz~(\ref{LIansatz}), 
which contains the single free quantity $m_{\rm ph}$, 
the corresponding ansatz for our Lorentz-violating model 
is parametrized by five free coefficients 
$\bar x$, $\bar y$, $m_{\rm ph}$, $m_c$, and $m_k$.

Paralleling the usual Lorentz-invariant reasoning,
we now use our ansatz~(\ref{LV-pole}) to replace 
$\slashed p$ and $p^2=\slashed p \slashed p$
by $\bar P$ 
in the expression for the two-point function~(\ref{LV-expansion1}).
To this end,
it is useful to write
\begin{equation}
p^2=\slashed p\slashed p=\bar P\bar P+2\bar m\bar P+\bar\beta\,,
\label{p-squared-bar-P}
\end{equation}
where at leading order in Lorentz violation
\begin{equation}
\bar\beta
=\bar m^2-2\bar x c^p_p-2\bar y\tilde k^p_p
\,.
\end{equation}
This produces the Lorentz-breaking analogue of Eq.~(\ref{LIansatzP}):
\begin{eqnarray}\label{LVansatzP}
\hspace{0em}\Gamma^{(2)}(\bar P) & = &
A(\bar P{}^{2}+2\bar m\bar P+\bar \beta,c^p_p,\tilde{k}^p_p)\bar P\nonumber\\
&&\hspace{-0em}{}+\bar m A(\bar P{}^{2}+2\bar m\bar P+\bar \beta,c^p_p,\tilde{k}^p_p)\openone\nonumber\\
&&\hspace{-0em}{}-M(\bar P{}^{2}+2\bar m\bar P+\bar \beta,c^p_p,\tilde{k}^p_p)\openone\nonumber\\&&
\hspace{-0em}{}+C(\bar P{}^{2}+2\bar m\bar P+\bar \beta)c^p_\gamma\nonumber\\
&&\hspace{-0em}{}-\bar x A(\bar P{}^{2}+2\bar m\bar P+\bar\beta,c^p_p,\tilde{k}^p_p)c^p_\gamma\nonumber\\
&&\hspace{-0em}{}+K(\bar P{}^{2}+2\bar m\bar P+\bar \beta)\tilde{k}^p_\gamma\nonumber\\
&&\hspace{-0em}{}-\bar y A(\bar P{}^{2}+2\bar m\bar P+\bar \beta,c^p_p,\tilde{k}^p_p)\tilde{k}^p_\gamma\,.
\end{eqnarray}
Although some higher-order terms in Lorentz violation appear in this expression for notational convenience, 
Eq.~(\ref{LVansatzP}) holds at linear order in $c^{\mu\nu}$ and $\tilde{k}^{\mu\nu}$.

We proceed by 
evaluating $\Gamma^{(2)}(\bar P)$ 
at $\bar P=0$. 
As $\bar P$ is our ansatz for the pole,
we must have $\Gamma^{(2)}(\bar P=0)=0$.
This yields
\begin{eqnarray}\label{LV_pole_eq}
0&=&
\big[
C(\bar \beta, c^p_p, \tilde{k}^p_p)
-\bar x A(\bar \beta, c^p_p, \tilde{k}^p_p)
\big]c^p_\gamma\nonumber\\
&&{}+\big[
K(\bar \beta, c^p_p, \tilde{k}^p_p)
-\bar y A(\bar \beta, c^p_p, \tilde{k}^p_p)
\big]\tilde{k}^p_\gamma\nonumber\\
&&{}+\big[
\bar m A(\bar \beta, c^p_p, \tilde{k}^p_p)
-M(\bar \beta, c^p_p, \tilde{k}^p_p)
\big]\openone\,.
\end{eqnarray}
Since $c^{\mu\nu}$ and $\tilde{k}^{\mu\nu}$ 
are in general not proportional~\cite{PropComment},
we take $c^p_\gamma$ and $\tilde{k}^p_\gamma$ 
to be linearly independent, 
so that each square bracket in Eq.~(\ref{LV_pole_eq}) 
must vanish separately. 
The two relations resulting from the first two brackets 
are needed at zeroth order in Lorentz violation;
they can be cast into the following form:
\begin{eqnarray}
\bar x &=&\frac{C(m_{\rm ph}^2)}{A(m_{\rm ph}^2)}
\,,
\label{x_eq}\\
\bar y &=&\frac{K(m_{\rm ph}^2)}{A(m_{\rm ph}^2)}
\,.
\label{y_eq}
\end{eqnarray}
Notice that $A(m_{\rm ph}^2,0,0)=A(m_{\rm ph}^2)$ 
is the conventional coefficient function for the Lorentz-invariant case.
In a similar manner, we have defined 
$C(m_{\rm ph}^2)\equiv C(m_{\rm ph}^2,0,0)$ and 
$K(m_{\rm ph}^2)\equiv K(m_{\rm ph}^2,0,0)$.
The relation arising from the third square bracket in Eq.~(\ref{LV_pole_eq})
is needed at first order in Lorentz violation, 
so we may expand in $c^p_p$ and $\tilde{k}^p_p$ as follows:
\begin{eqnarray}\label{fullM_eq}
{}\hspace{-1.8em} 0 \!&=&\!
\bar m A(\bar \beta, c^p_p, \tilde{k}^p_p)
-M(\bar \beta, c^p_p, \tilde{k}^p_p)\nonumber\\
{}\hspace{-1.8em} &=&\!\big[
m_{\rm ph}A(m_{\rm ph}^2)-M(m_{\rm ph}^2)
\big]\nonumber\\
{}\hspace{-1.8em}&&\!{}+\partial_{c^p_p}
\big[
\bar m A(\bar \beta, c^p_p, \tilde{k}^p_p)
-M(\bar \beta, c^p_p, \tilde{k}^p_p)
\big]_{c^p_p,\tilde{k}^p_p=0}\;c^p_p\nonumber\\
{}\hspace{-1.8em}&&\!{}+\partial_{\tilde{k}^p_p}
\big[
\bar m A(\bar \beta, c^p_p, \tilde{k}^p_p)
-M(\bar \beta, c^p_p, \tilde{k}^p_p)
\big]_{c^p_p,\tilde{k}^p_p=0}\;\tilde{k}^p_p\,.
\end{eqnarray}
We note that 
$M(m_{\rm ph}^2,0,0)=M(m_{\rm ph}^2)$ 
is the usual Lorentz-symmetric coefficient function.
As we have taken $c^{\mu\nu}$ and $\tilde{k}^{\mu\nu}$
to be linearly independent,
each square bracket in Eq.~(\ref{fullM_eq}) must 
be equal to zero individually, 
which yields three algebraic equations:
\begin{eqnarray}
0&=&m_{\rm ph}A(m_{\rm ph}^2)-M(m_{\rm ph}^2)\,,
\label{m1_eq}\\
0&=&\partial_{c^p_p}
\big[
\bar m A(\bar \beta, c^p_p, \tilde{k}^p_p)
-M(\bar \beta, c^p_p, \tilde{k}^p_p)
\big]_{c^p_p,\tilde{k}^p_p=0}\,,
\label{m2_eq}\\
0&=&\partial_{\tilde{k}^p_p}
\big[
\bar m A(\bar \beta, c^p_p, \tilde{k}^p_p)
-M(\bar \beta, c^p_p, \tilde{k}^p_p)
\big]_{c^p_p,\tilde{k}^p_p=0}\,.
\label{m3_eq}
\end{eqnarray}
The five relations~(\ref{x_eq}), (\ref{y_eq}), (\ref{m1_eq}), (\ref{m2_eq}), and~(\ref{m3_eq}) 
determine the five parameters 
$m_{\rm ph}$, $\bar x$, $\bar y$, $m_c$, and $m_k$
in our ansatz for the pole~(\ref{LV-pole}) 
in terms of the functions $A$, $C$, $K$, and $M$, 
which are calculable in perturbation theory. 
It follows that the expression for $\bar P$ 
is now completely fixed.
These five relations constitute a direct generalization of 
Eq.~(\ref{LIdetermination})
valid in the usual Lorentz-symmetric context. 
In particular, 
Eq.~(\ref{LIdetermination}) governing the Lorentz-invariant case 
is identical to Eq.~(\ref{m1_eq}) in the Lorentz-breaking situation.
We remark that 
as a consequence
the value of the physical mass $m_{\rm ph}$---which we interpret as the momentum-independent piece of the coefficient of $\overline{\psi}\psi$---remains 
unaffected by Lorentz violation.

The remaining task is to extract the field-strength renormalization 
${\cal Z}_R$. 
To this end, 
we may again proceed in a manner similar to the Lorentz-invariant situation
and expand the perturbation-theory two-point function 
$\Gamma^{(2)}(\bar P)$ 
around $\bar P=0$.
As opposed to the conventional case,
where only a single nontrivial matrix given by $\slashed p$ appears,
the present Lorentz-violating situation 
involves the three matrices 
$\slashed p$, $c^p_\gamma$, and $\tilde{k}^p_\gamma$,
which are in general noncommuting.
For this reason, 
the expansion of $\Gamma^{(2)}(\bar P)$, 
which we have relegated to Appendix~\ref{orderingApp},
requires special care
to avoid matrix-ordering ambiguities.
We find
\begin{eqnarray}\label{genZ_R}
{\cal Z}_R^{-1}\!&=&\!A(\bar \beta,c^p_p,\tilde{k}^p_p)
+2\bar m\big[\bar m A'(\bar \beta,c^p_p,\tilde{k}^p_p)-M'(\bar \beta,c^p_p,\tilde{k}^p_p)\big]\nonumber\\
&&{}\!+2\big[C'(m_{\rm ph}^2)-\bar x A'(m_{\rm ph}^2)\big]c^p_p\nonumber\\
&&{}\!+2\big[K'(m_{\rm ph}^2)-\bar y A'(m_{\rm ph}^2)\big]\tilde k^p_p\,,
\label{LV-Z}
\end{eqnarray}
where a prime denotes the derivative with respect to the first argument.


\section{One-loop calculation of the modified propagator}
\label{one-loop}

\begin{center}
\begin{figure}[h]
\includegraphics[width=0.75\hsize]{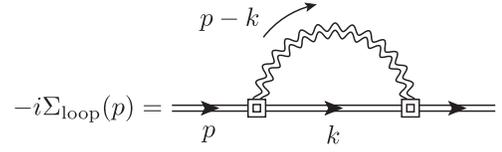}
\caption{One-loop Feynman diagram 
for the determination of the fermion self-energy 
at ${\cal O}(\alpha)$. 
The complete $\Sigma$ at this order 
also contains counterterm insertion diagrams 
that have not been included above. 
}
\label{Sigma-loop}
\vskip-10pt
\end{figure}
\end{center}

An interesting question concerns
the determination of the functions
$f_0(p^2)$,
$f_1(p^2)$,
$f_2^c(p^2)$,
$f_2^{\tilde k}(p^2)$,
$f_3^c(p^2)$,
$f_4^c(p^2)$,
$f_3^{\tilde k}(p^2)$, and
$f_4^{\tilde k}(p^2)$
perturbatively at leading order in the fine-structure constant $\alpha$. 
To this end, 
we will adopt the perturbative scheme based on the expressions
(\ref{model-lagrangian-split})--(\ref{L2}),
in which the propagators are built from the full quadratic Lagrange density,
including the Lorentz-violating parts.
The corresponding Feynman rules are presented in Appendix~\ref{2pointrules},
and the only loop diagram involved in
the fermion self-energy is shown in Fig.~\ref{Sigma-loop}.

This diagram together with the corresponding counterterm contributions 
contains the conventional Lorentz-symmetric $\mathcal{O}(\alpha)$ results
\begin{align}
\label{f_0}
f_0(p^2)
&=\frac\alpha\pi
\left[-\frac12-\gamma_E-\int_0^1 dy
\,\ln\left(\frac\Delta{4\pi\mu^2}\right)\right] \,,
\\
f_1(p^2)
&=\frac{\alpha}{4\pi}
\left[1+\gamma_E+2\int_0^1 dy
\,(1-y)\ln\left(\frac\Delta{4\pi\mu^2}\right)\right]
\label{f_1}
\end{align}
Here,
\begin{equation}
\Delta=-y(1-y)p^2+y(m^2-m_\gamma^2)+m_\gamma^2,
\end{equation}
$\gamma_E=0.57721\ldots$ denotes the Euler--Mascheroni constant, 
$m_\gamma$ represents a fictitious photon mass introduced as an infrared regulator,
and the arbitrary mass scale $\mu$ is a remnant from dimensional regularization. 
The integrations over 
$y$ 
can be performed requiring 
$p^2$ to be close to the conventional mass shell:
$(m-m_\gamma)^2<p^2<(m+m_\gamma)^2$.
They yield the infrared-finite limits on the (conventional) mass shell,
\begin{eqnarray}
f_0(m^2)&=&\frac\alpha\pi\left[\frac32-\gamma_E
-\ln\left(\frac{m^2}{4\pi\mu^2}\right)\right]\,,\\
f_1(m^2)&=&\frac\alpha\pi\left[-\frac12+\frac{\gamma_E}4+
\frac14\ln\left(\frac{m^2}{4\pi\mu^2}\right)\right].
\end{eqnarray}
For the on-shell values of their first and second derivatives we find 
at leading order in $m_\gamma$,
\begin{eqnarray}
f_0'(m^2)&=&\frac\alpha{\pi m^2}\left[\ln\left(\frac m{m_\gamma}\right)-1\right],\\
f_0''(m^2)&=&\frac\alpha{4\pi m^4}\left[
\pi\,\frac m{m_\gamma}
-8\ln\left(\frac m{m_\gamma}\right)
+6
\right],\\
f_1'(m^2)&=&\frac\alpha{\pi m^2}\left[-\frac12\ln\left(\frac m{m_\gamma}\right)+\frac34\right],\\
f_1''(m^2)&=&\frac\alpha{8\pi m^4}\left[
-\pi\,\frac m{m_\gamma}
+12\ln\left(\frac m{m_\gamma}\right)
-14
\right].\quad
\end{eqnarray}

\vskip-20pt
An explicit calculation also yields ultraviolet-finite expressions for 
the remaining, Lorentz-violating contributions to $\Sigma$.
For the functions $f_i^c(p^2)$, 
we obtain
\begin{eqnarray}
{}\hspace{-6mm}f_2^c(p^2)
&=&\frac{\alpha}{2\pi}\biggl[\frac16-\frac{5\gamma_E}6\nonumber\\
&&\>{}-\int_0^1dy(1-y)(1+2y)\ln\left(\frac\Delta{4\pi\mu^2}\right)\biggl],
\label{f_2^c(p^2)}\\
{}\hspace{-6mm}f_3^c(p^2)
&=&\frac{2\alpha m^2}\pi \int_0^1dy\frac{y(1-y)^2}\Delta\,,\\
{}\hspace{-6mm}f_4^c(p^2)
&=&-\frac{\alpha m^2}\pi \int_0^1dy\frac{y(1-y)^3}\Delta\,.
\label{f_4^c(p^2)}
\end{eqnarray}
While $f_2^c(p^2)$ is infrared finite, 
this is not the case for $f_3^c(p^2)$ and $f_4^c(p^2)$: 
the latter both diverge on the conventional mass shell $p^2=m^2$ 
when the limit $m_\gamma\to0$ is taken:
\begin{eqnarray}
f_2^c(m^2)
&=&\frac\alpha\pi\left[\frac{10}9-\frac{5\gamma_E}{12}
-\frac5{12}\ln\left(\frac{m^2}{4\pi\mu^2}\right)\right],
\label{f_2^c(m^2)}\\
f_3^c(m^2)
&=&\frac\alpha\pi\left[2\ln\left(\frac m{m_\gamma}\right)-3\right],\\
f_4^c(m^2)
&=&\frac\alpha\pi\left[-\ln\left(\frac m{m_\gamma}\right)+\frac{11}6\right].
\label{f_4^c(m^2)}
\end{eqnarray}

Below, 
we will also need the derivatives of these functions at their
on-shell value $p^2=m^2$.
To this effect, we use that
\begin{equation}
\frac d{dp^2}\ln\Delta=-\frac{y(1-y)}\Delta\,, \qquad
\frac d{dp^2}\frac1\Delta=\frac{y(1-y)}{\Delta^2}\,.
\label{derivative-Delta}
\end{equation}
Applying these relations to Eqs.~(\ref{f_2^c(p^2)})--(\ref{f_4^c(p^2)}), 
and evaluating the integrals on the conventional mass shell 
yields  the following results:
\begin{eqnarray}
\hspace{-5mm}f_2^c{}'(m^2)
&=&\frac\alpha{2\pi m^2}\left[\ln\left(\frac m{m_\gamma}\right)
-\frac56\right],\\
\hspace{-5mm}f^c_3{}'(m^2)
&=&\frac\alpha{\pi m^2}\left[\frac\pi2\frac m{m_\gamma}
-6\ln\left(\frac m{m_\gamma}\right)+7\right],\\
\hspace{-5mm}f_4^c{}'(m^2)
&=&\frac\alpha{\pi m^2}\left[-\frac\pi4\frac m{m_\gamma}
+4\ln\left(\frac m{m_\gamma}\right)-\frac{35}6\right].
\end{eqnarray}
Note that $f_3^c{}'(p^2)$ and $f_4^c{}'(p^2)$ have linear,
rather than logarithmic, infrared divergences.

For the coefficients $f_i^{\tilde k}(p^2)$ one finds
\begin{eqnarray}
f_2^{\tilde k}(p^2)
&=&\frac{\alpha}\pi\Bigl[\frac1{12}+\frac{\gamma_E}3\nonumber\\
&&{}+\frac12\int_0^1dy
(1-y^2)\ln\left(\frac\Delta{4\pi\mu^2}\Bigr) \right]\,,
\label{f_2^k(p^2)}\\
f_3^{\tilde k}(p^2)
&=&\frac{\alpha m^2}{\pi} \int_0^1dy\frac{y^2(1-y)}\Delta\,,\\
f_4^{\tilde k}(p^2)
&=&-\frac{\alpha m^2}{2\pi} \int_0^1dy\frac{y^2(1-y)^2}\Delta\,.
\label{f_4^k(p^2)}
\end{eqnarray}
All three coefficients $f_i^{\tilde k}(p^2)$ are infrared finite on the conventional mass shell:
\begin{eqnarray}
f_2^{\tilde k}(m^2)&=&\frac\alpha\pi\left[-\frac{29}{36}+\frac{\gamma_E}3
+\frac13\ln\left(\frac{m^2}{4\pi\mu^2}\right)\right]\,,
\label{f_2^k(m^2)}\\
f_3^{\tilde k}(m^2)&=&\frac\alpha{2\pi}\,,\\
f_4^{\tilde k}(m^2)&=&-\frac\alpha{6\pi}\,.
\label{f_4^k(m^2)}
\end{eqnarray}
Applying the relations~(\ref{derivative-Delta}) to Eqs.~(\ref{f_2^k(p^2)})--(\ref{f_4^k(p^2)})
and evaluating the integrals at $p^2=m^2$, 
one finds the infrared-divergent expressions 
\begin{eqnarray}
f_2^{\tilde k}{}'(m^2)
&=&\frac\alpha{\pi m^2}\left[-\frac12\ln\left(\frac m{m_\gamma}\right)+\frac7{12}\right],\\
f_3^{\tilde k}{}'(m^2)
&=&\frac\alpha{\pi m^2}\left[\ln\left(\frac m{m_\gamma}\right)-2\right],\\
f_4^{\tilde k}{}'(m^2)
&=&\frac\alpha{\pi m^2}\left[
-\frac12\ln\left(\frac m{m_\gamma}\right)+\frac76\right].
\end{eqnarray}

Let us now see how the above one-loop calculation 
and the general formalism developed in Sec.~\ref{2pointfunction}
can be used to determine explicit ${\cal O}(\alpha)$ expressions for 
$\bar P$ and ${\cal Z}_R$ 
in terms of our model's coefficients.
This section's results 
together with the definitions~(\ref{ACKM_Def})
yield the one-loop approximation of Eq.~(\ref{m1_eq}),
which can be solved at ${\cal O}(\alpha)$:
\begin{eqnarray}\label{phys_mass}
\hspace{-5mm}m_{\rm ph}\!&=&\!m\big[1+f_0(m^2)+f_1(m^2)\big]\nonumber\\
&=&\!m+\frac{\alpha}{\pi}\left[1-\frac{3\gamma_E}4-
\frac34\ln\left(\frac{m^2}{4\pi\mu^2}\right)\right]m\,.
\end{eqnarray}
With this result,
the parameters $\bar x$ and $\bar y$ directly follow 
from Eqs.~(\ref{x_eq}) and~(\ref{y_eq}) at one-loop order:
\begin{eqnarray}
\bar x &=&1+f_1(m^2)-f_2^c(m^2)\nonumber\\
&=&1+2\frac{\alpha}{\pi}\left[-\frac{29}{36}+\frac{\gamma_E}{3}
+\frac13\ln\left(\frac{m^2}{4\pi\mu^2}\right)\right],
\label{x0}\\
\bar y &=&-f_2^{\tilde k}(m^2)\nonumber\\
&=&-\frac{\alpha}{\pi}\left[-\frac{29}{36}+\frac{\gamma_E}{3}
+\frac13\ln\left(\frac{m^2}{4\pi\mu^2}\right)\right].
\label{y0}
\end{eqnarray}
We continue with the determination of $m_c$ and $m_k$ 
from Eqs.~(\ref{m2_eq}) and~(\ref{m3_eq}), 
respectively.
To this end, 
notice that $\bar \beta =m^2-2c^p_p+\mathcal{O}(\alpha)$, 
which gives
\begin{equation}
\big[\partial_{c^p_p}f_0(\bar \beta)\big]_{c^p_p,\tilde{k}^p_p=0}
=-2f'_0(m^2)+\mathcal{O}(\alpha^2)\,,
\label{Taylor-f1}
\end{equation}
with an analogous result for the function $f_1$.
We then find at leading order in $\alpha$:
\begin{eqnarray}
m_c\!&=&\!-2m\big[f_0'(m^2)+f_1'(m^2)\big]+
\frac{1}{m}\big[f_3^c(m^2)+f_4^c(m^2)\big]\nonumber\\
&=&\!-\frac{2\alpha}{3\pi m}\,,
\label{m10}\\
m_k\!&=&\!\frac{1}{m}
\big[f_3^{\tilde k}(m^2)+f_4^{\tilde k}(m^2)\big]\nonumber\\
&=&\!\frac{\alpha}{3\pi m}\,.
\label{m01}
\end{eqnarray}

The above results together with the general formula~\rf{genZ_R} 
allow us to give the following explicit form 
of the wave-function renormalization at one-loop order:
\begin{align}
\mathcal{Z}_R^{-1}&
=1-f_1(m^2)-2m^2\Big[f_0'(m^2)+f_1'(m^2)\Big]\nonumber\\
&\quad{}+2c^p_p\bigg[2f_1'(m^2)+2m^2f_0''(m^2)+2m^2f_1''(m^2)
\nonumber\\
&\quad{}-f_2^c{}'(m^2)-f_3^c{}'(m^2)
-f_4^c{}'(m^2)-\frac{f_4^c(m^2)}{2 m^2}\bigg]\nonumber\\
&\quad{}-2\tilde k^p_p\bigg[f_2^{\tilde k}{}'(m^2)+f_3^{\tilde k}{}'(m^2)+f_4^{\tilde k}{}'(m^2)
+\frac{f_4^{\tilde k}(m^2)}{2m^2}\bigg]\nonumber\\
&= 1-\frac\alpha\pi\left[\ln\left(\frac{m}{m_\gamma}\right)-1+\frac{\gamma_E}{4}
+\frac14\ln\left(\frac{m^2}{4\pi\mu^2}\right)\right]\nonumber\\
&\quad{}-\frac{2\alpha}{3\pi m^2}(2\,c^p_p-\tilde k^p_p)\,.
\label{Z_R-2}
\end{align}
The Lorentz-symmetric piece of $\mathcal{Z}_R$
is identical to the conventional one-loop wave-function renormalization constant
for the fermion in QED.
In particular, 
it exhibits the usual logarithmic infrared divergence.
On the other hand, 
the linear and logarithmic infrared divergences that 
are present in the various individual $f$ coefficient functions
that multiply $c^p_p$ and $\tilde k^p_p$
in the intermediate step
are absent in the final expression of Eq.~(\ref{Z_R-2}).
As is well known, 
the Lorentz-symmetric infrared divergences cancel in scattering cross sections
when the contributions of soft-photon emission 
from the corresponding external legs
are taken into account.
In Sec.~\ref{Coulomb-scattering}, 
we will verify that 
these soft-photon contributions 
do not introduce additional infrared divergences 
proportional to $c^p_p$ or $\tilde k^p_p$, 
so that physical observables remain infrared finite. 
We also note that
the Lorentz-violating radiative corrections 
indeed appear in the combination $(2c^{\mu\nu}-\tilde k^{\mu\nu})$, 
as anticipated in Sec.~\ref{model-basics}.

A novel feature relative to the Lorentz-invariant situation 
is the momentum dependence of ${\cal Z}_R$.
We do not see any conceptual issues arising from this feature 
in the momentum range that is of phenomenological interest.
However, 
bounds on ${\cal Z}_R$ obtained nonperturbatively on physical grounds 
may perhaps be used together with Eq.~\rf{Z_R-2} 
to investigate the validity of perturbation theory 
at ultrahigh momenta.

We are now also in a position to determine explicit expressions 
for physically measurable model parameters.
These can be identified via inspection of 
Eqs.~\rf{LV-pole} and~\rf{m-first-order}. 
For example, 
it is apparent that $m_{\rm ph}$, 
which is given explicitly at order $\alpha$ in Eq.~\rf{phys_mass},
is the measurable mass parameter 
governing the propagation of asymptotic states.
The terms 
$\bar x c^p_\gamma$ and $\bar y k^p_\gamma$ 
in the inverse propagator~\rf{LV-pole}
cannot individually be resolved in an experiment. 
Instead, 
their sum is interpreted to determine 
the physical $c$ coefficient denoted by $c_{\rm ph}^{\mu\nu}$,
\begin{align}
c_{\rm ph}^{\mu\nu}&=\bar x c^{\mu\nu}+\bar y \tilde{k}^{\mu\nu}
\nonumber\\
&= c^{\mu\nu}
-\frac{\alpha}{3\pi}\left[\frac{29}{12}-\gamma_E
-\ln\left(\frac{m^2}{4\pi\mu^2}\right)\right]
(2c^{\mu\nu}-\tilde k^{\mu\nu})\,,
\label{c-phys}
\end{align}
where the last equality is based on our above one-loop results.
Note again that the radiative corrections depend on
$(2c^{\mu\nu}-\tilde k^{\mu\nu})$
in line with our discussion in Sec.~\ref{model-basics}.
Note also that 
$c_{\rm ph}^{\mu\nu}$ is free of infrared divergences.
We finally remark that unlike $c_{\rm ph}^{\mu\nu}$,
both $c^{\mu\nu}$ and the scale $\mu$ are unphysical
renormalization-scheme-dependent quantities. 
In particular,
the running of $c^{\mu\nu}$ 
with $\mu$~\cite{Kostelecky:2001jc}
cancels the explicit appearances of $\mu$ 
in Eq.~\rf{c-phys},
so that $c_{\rm ph}^{\mu\nu}$ is independent of $\mu$.
This situation is completely analogous to 
the conventional relation~(\ref{phys_mass})
expressing the constant $m_{\rm ph}$ in terms of $\mu$ 
and the running mass $m$.

Relative to the tree-level expression 
\begin{equation}
\bar P_{\scriptsize\rm tree}(p)=\slashed p + c^p_\gamma - m\,,
\label{P-tree}
\end{equation}
the full Dirac operator~\rf{LV-pole} 
not only contains the above radiative corrections to 
the existing $m$ and $c^{\mu\nu}$ pieces, 
but it also displays the new structures 
$m_c c^p_p$ and $m_k k^p_p$,
which possess more than a single power of momentum.
Our results~\rf{m10} and~\rf{m01} show that
the coefficients $m_c$ and $m_k$ are nonzero.
At one-loop order, 
the asymptotic Dirac operator $\bar P_1$ 
can therefore be written in the form~\cite{k_comment}
\newcommand{\phy}{{\scriptsize\rm ph}}
\begin{equation}
\bar P_1=\slashed p + (c_\phy)^p_\gamma - m_\phy +
\frac{\alpha}{3\pi m}\big[2(c_\phy)^p_p-(\tilde k_\phy)^p_p\big]\,.
\label{P-phys}
\end{equation}
In accordance with the general expectation
discussed in Sec.~\ref{model-basics}, 
the new structures---shown as the last term in Eq.~\rf{P-phys}---enter 
in the combination $(2c^{\mu\nu}-\tilde k^{\mu\nu})$.
The fact that in addition to 
{\em shifts in existing model parameters}, 
asymptotic states in Lorentz-violating field theories 
also acquire novel {\em higher-derivative $\mu$-independent structures} 
represents a key finding of our work.
We remark that
this result is fully compatible with the general
form of the one-particle fermion pole of the K\"all\'en--Lehmann
representation derived in Ref.~\cite{Potting:2011yj}.

With the asymptotic Dirac operator~\rf{LV-pole} 
and its one-loop approximation~\rf{P-phys} at hand, 
we may determine various properties of the external fermion states. 
For example, 
the corresponding two forms of the dispersion relation 
are given by
\begin{align}
0&=p^2+2\bar x c^p_p+2\bar y\tilde k^p_p-m_{\rm ph}^2-2m_{\rm ph} (m_c c^p_p+m_k k^p_p) \nonumber\\
&=p^2+2(c_{\rm ph})^p_p-m_{\rm ph}^2
+\frac{2\alpha}{3\pi }\big[2(c_\phy)^p_p-(\tilde k_\phy)^p_p\big].
\label{dispersion-relation}
\end{align}
The associated eigenspinors and some of their properties 
will be discussed as part of Sec.~\ref{external-LSZ}.

We mention again the pleasing fact that
the procedure outlined in this section
consistently avoids infrared divergences in physical observables,
despite their presence in most of the $f$ coefficient functions
and their derivatives.
For example, 
we see that 
the dispersion relation~(\ref{dispersion-relation}) 
is infrared finite. 
One can also readily verify that
this is in fact the case for all coefficients 
$A(\bar\beta,c^p_p,k^p_p)$, $M(\bar\beta,c^p_p,k^p_p)$, 
$C(\bar\beta,c^p_p,k^p_p)$, and $K(\bar\beta,c^p_p,k^p_p)$ 
up the order required for our purposes
and also for the coefficients $\bar x$, $\bar y$, $m_c$, and $m_k$.
We conjecture that 
infrared divergences will continue to cancel at any order in the
perturbative expansion.

Passing the momenta to derivatives, 
it becomes apparent that 
the Lagrangian describing the asymptotic on-shell free fermion field
acquires higher spacetime derivatives.
This is the case for both the Lagrangian derived 
from the original proper two-point function 
and for the one derived from the operator $\bar P (p)$.
Often, 
it might be possible to avoid working directly with 
the physical field
and employ the bare field instead, 
treating the higher-dimensional Lorentz-violating terms perturbatively. 
However, 
this becomes problematic or impossible if we consider on-shell external
states, 
as in the derivation of the LSZ
reduction formula,
which computes scattering amplitudes for on-shell external physical states.
In the next section, 
we will analyze how this situation generalizes
to loop-corrected Lorentz-violating Lagrangians.


\section{External states in Feynman diagrams and the LSZ formula}
\label{external-LSZ}

How do the Lorentz-violating radiative corrections 
parametrized by the coefficient functions $\bar x$, $\bar y$, and $\bar m$ 
in Eq.~(\ref{LV-pole}) contribute to S-matrix elements? 
Let us reflect a moment on how we can determine the latter.

In the quantum-field description of scattering experiments,
it is presupposed that 
the Fock space of physical states is generated from 
a unique vacuum by free fields $\psi_{in}(x)$ and $\bar\psi_{in}(x)$ 
(here we will only consider fermions in the asymptotic states 
and ignore the possibility of photons).
One assumes that 
the coupling terms in the equations of motion 
are affected by some adiabatic cut-off function 
equal to unity at finite times
and vanishing smoothly as $|t|\to \infty$, 
and the particles in the initial and final states have become well separated.
Then, 
according to the usual adiabatic hypothesis 
the interacting fields $\psi(x)$ and $\bar\psi(x)$ are presumed to satisfy,
in a weak sense,
\begin{equation}
\psi(x)\to Z^{1/2}\psi_{in}(x)\qquad\textrm{as}\qquad t\to-\infty\qquad
\label{LI-Z}
\end{equation}
[and similarly for $\bar\psi(x)$]
for some normalization constant $Z$ that 
should be smaller than one,
in order to account for the fact that 
the content of the state $\psi(x)|0\rangle$ 
is not exhausted by the matrix elements
with one-particle states, 
while $\psi_{in}|0\rangle$ is.

In Sec.~\ref{one-loop}, 
we saw that 
for the Lorentz-violating model we are considering 
the normalization constant analogous to
the constant $Z$ in Eq.~(\ref{LI-Z}) is not only Lorentz violating,
but becomes dependent on the momentum of the external particle:
$Z\to\mathcal{Z}_R(p)$, see Eq.~(\ref{Z_R-2}).
To see how this will affect the usual treatment of external states 
in scattering amplitudes,
let us begin by 
looking at the free field $\psi_{in}(x)$ 
(the out-field $\psi_{out}(x)$ will be analogous).
Consider the spinor wave functions
$u^s_{in}(\vec p)$ and $v^s_{in}(\vec p)$ 
of the physical field $\psi_{in}(x)$.
They are modified with respect to the Lorentz-invariant situation.
While in the latter case we have
$(\slashed p-m)u^s_{in}(\vec p)=0$ (with $p^0=\omega_p>0$)
and $(\slashed p-m)v^s_{in}(\vec p)=0$ (with $p^0=-\omega_p<0$),
the spinors now satisfy
\begin{equation}
\label{modified-u-eom}
\bar P(p)u^s_{in}(\vec p)=0\,,\qquad (p^0>0)
\end{equation}
for the positive-energy solutions and
\begin{equation}
\bar P(p)v^s_{in}(\vec p)=0\,,\qquad (p^0<0)
\label{modified-v-eom}
\end{equation}
for the negative-energy solutions corresponding to a given 3-momentum.
Thus, 
we conclude that 
our model's external spinors, 
unlike in the Lorentz-invariant case,
are modified by the one-loop radiative-correction
terms calculated in the previous section.

Note also that 
we get the Lorentz-violating multiplicative contribution
$\mathcal{Z}_R\bigl((c,k)^p_p\bigr)$ 
(wave-function renormalization)
to the S-matrix for every external fermion that
factors out of the fermion propagator pole. 

Let us analyze in some more detail 
the new equations of motion for the spinors, 
Eqs.~(\ref{modified-u-eom}) and~(\ref{modified-v-eom}).
To all orders in $\alpha$
and to first order in Lorentz-violating parameters, 
we can use Eqs.~(\ref{LV-pole}) 
and~(\ref{m-first-order}),
which yield
\begin{equation}
\left(\slashed p+\bar x \,c^p_\gamma+\bar y \,\tilde k^p_\gamma
-m_{\rm ph}-m_c\,c^p_p-m_k\,\tilde k^p_p\right)u^s_{in}(\vec p)=0\,.
\label{eom_u_quadratic}
\end{equation}
For fixed 3-momentum $\vec p$, 
the value of $p^0$ in Eq.~(\ref{eom_u_quadratic})
is determined as the positive root of the dispersion relation~(\ref{dispersion-relation}).
Every term in the dispersion relation is of even order
in the 4-momentum, 
so that when $(p^0,\vec p\,)$, 
where $p^0>0$, 
satisfies the dispersion relation~(\ref{dispersion-relation}), 
so does $(-p^0,-\vec p\,)$.
The latter solution is taken to correspond to $v^s_{in}(\vec p)$,
an antifermion with momentum $\vec p$ and energy $p^0$.
Thus,
\begin{equation}
\left(\slashed p+\bar x \,c^p_\gamma+\bar y \,\tilde k^p_\gamma
+m_{\rm ph}+m_c\,c^p_p+m_k\,\tilde k^p_p\right)v^s_{in}(\vec p)=0\,,
\label{eom_v_quadratic}
\end{equation}
where $p^0$ takes the same value as in Eq.~(\ref{eom_u_quadratic}).
The fact that 
a fermion and an antifermion with the same momentum
have equal energy 
is a consequence of CPT invariance, 
which is unbroken by the $c^{\mu\nu}$ 
(and by the $\tilde k^{\mu\nu}$) coefficients.

On the other hand, 
in general there are Lorentz-violating quadratic terms 
in the dispersion relation~(\ref{dispersion-relation}) that 
mix $p^0$ and $\vec p$.
As a consequence, 
the fact that $(p^0,\vec p\,)$ (with $p^0>0$) 
satisfies Eq.~(\ref{dispersion-relation}) 
does not imply the same for $(-p^0,\vec p\,)$.
This expresses the fact that 
the $c^{0i}$ (and $\tilde k^{0i}$) violate parity.
Another useful observation is that 
the dispersion relation~(\ref{dispersion-relation}) is not sensitive to the spin label $s$.
Note that spin-dependence does play a role for some of the other
types of SME coefficients, 
but we will not consider them in this work.

We see from Eqs.~(\ref{eom_u_quadratic}) and~(\ref{eom_v_quadratic}) that
the equation of motion has terms quadratic
in the momentum due to the presence of $c^p_p$ and $\tilde k^p_p$.
These terms
make a rigorous analysis of the equation of motion for
the external fermion field and a quantization of the latter
along the lines of Appendix~\ref{canonical} problematic.
For instance, 
they likely introduce spurious unphysical solutions.
For this reason, 
we use the zeroth-order dispersion relation 
$p^0=\sqrt{\vec p\,^2+m^2}\equiv \omega_p$
to substitute for $c^p_p$,
\begin{equation}
c^p_p\to c^{00}\omega_p^2-2c^{0i}p^0p^i+c^{ij}p^ip^j
\end{equation}
and similarly for $\tilde k^p_p$.
For simplicity, 
we will suppress the $\tilde k$ terms in the following.
Equation~(\ref{eom_u_quadratic}) becomes
\begin{align}
\bigl(\Gamma^\mu p_\mu&-m_{\rm ph}-m_cc^{00}\omega_p^2
\nonumber\\
&{}+2m_cc^{0i}p^0p^i-m_cc^{ij}p^ip^j\bigr)u^s_{in,1}(\vec p)=0\,,
\label{eom_u_linearized}
\end{align}
with
\begin{equation}
\Gamma^\mu=\gamma^\mu+\bar x c^{\mu\nu}\gamma_\nu\,.
\end{equation}
The spinor $u^s_{in,1}(\vec p)$ satisfying Eq.~(\ref{eom_u_linearized})
differs from the original one $u^s_{in}(\vec p)$ by terms of second
order (and higher) in the Lorentz-violating coefficients.
We remark that
this higher-order difference between the spinors 
permits a self-consistent treatment of the asymptotic Hilbert space in terms of $u^s_{in,1}(\vec p)$,
while also allowing us to switch back to the original spinors $u^s_{in}(\vec p)$
at a later point in the calculation.

With these considerations, 
we can proceed as in Appendix~\ref{canonical}.
The equation of motion can be written in the form of an
eigenvalue equation:
\begin{align}
\tilde\Gamma^0(\vec p\,)^{-1}\Bigl[\Gamma^ip^i+m_{\rm ph}+m_cc^{00}\omega_p^2
+&m_cc^{ij}p^ip^j\Bigr]u^s_{in,1}(\vec p)\nonumber\\
&{}=p^0u^s_{in,1}(\vec p)\,,
\label{eigenvalue-equation-Gamma0}
\end{align}
where
\begin{equation}
\tilde\Gamma^0(\vec p\,)=\Gamma^0+2m_cc^{0i}p^i\,.
\end{equation}
The operator acting on the left-hand side of Eq.~(\ref{eigenvalue-equation-Gamma0}) on
the spinor is Hermitian with respect to the inner product
\begin{equation}
(u_1|u_2) \equiv \bar u_1\tilde\Gamma^0(\vec p\,)u_2\,, 
\end{equation}
which is different from that in Eq.~(\ref{innerproduct}). 
Consequently, it has real eigenvalues, with the corresponding eigenspinors
forming an orthonormal basis in spinor space 
$\{u^{s=1}_{in,1}(\vec p), u^{s=2}_{in,1}(\vec p),
v^{s=1}_{in,1}(-\vec p),v^{s=2}_{in,1}(-\vec p)\}$ 
satisfying the relations
\begin{align}
\bar u^r_{in,1}(\vec p)\tilde\Gamma^0(\vec p\,) u^s_{in,1}(\vec p)=
\frac{\omega_p}{m}\delta_{rs}\,,\nonumber\\
\bar v^r_{in,1}(-\vec p)\tilde\Gamma^0(\vec p\,) v^s_{in,1}(-\vec p)=
\frac{\omega_p}{m}\delta_{rs}\,,
\label{order1-norm-u,v}
\end{align}
as well as
\begin{equation}
\label{order1-is-matrix}
\sum_{s=1}^2\big[
u^s_{in,1}(\vec p)\bar u^s_{in,1}(\vec p)
+v^s_{in,1}(-\vec p)\bar v^s_{in,1}(-\vec p)
\big]\tilde\Gamma^0(\vec p\,)=\frac{\omega_p}{m}\openone\,,
\end{equation}
in analogy to 
Eqs.~(\ref{norm-u,v}) and~(\ref{is-matrix}).
Incidentally, 
note the Hermiticity relation 
$\tilde\Gamma^0(\vec p\,)^\dagger=\gamma^0\tilde\Gamma^0(\vec p\,)\gamma^0$ for $\tilde\Gamma^0(\vec p\,)$.

The free field has the Fourier decomposition
\begin{align}
\psi_{in}(x)=&\int\!\frac{d^3\vec{p}}{(2\pi)^3}\,\frac{m}{\omega_p}
\sum_{s=1}^2\big[
b^{in}_s(\vec p)u^{s}_{in,1}(\vec p)e^{-ip\cdot x}
\nonumber\\
&
\hspace{7em}+d_s^{in}{}^\dagger(\vec p)v^s_{in,1}(\vec p)e^{ip\cdot x}\big]\,.
\label{FmodeExpansion}
\end{align}
From Eq.~(\ref{eom_u_linearized}) we see that 
it satisfies the (linearized) equation of motion,
\begin{align}
\bigl[i\Gamma^\mu \partial_\mu&-m_{\rm ph}+m_cc^{00}(\nabla^2-m^2)
\nonumber\\
&{}
-2m_cc^{0i}\partial^0\partial^i+
m_cc^{ij}\partial_i\partial_j\bigr]\psi_{in}(x)=0\,.
\label{eom-psi_in}
\end{align}
The creation and annihilation operators can be expressed by 
the following projections,
\begin{eqnarray}
\label{order1-oscillator-expression-b}
b^{in}_s{}^\dagger(\vec p) \!\! &=& \!\!\!
\int \!\! d^3x\,e^{-i p\cdot x}\bar\psi_{in}(x)\tilde\Gamma^0(\vec p\,)u^s_{in,1}(\vec p)\,,\\
d^{in}_s{}^\dagger(\vec p) \!\! &=&\!\!\!
\int \!\! d^3x\,e^{-i p\cdot x}\bar v^s_{in,1}(\vec p)\tilde\Gamma^0(-\vec p\,)\psi_{in}(x)\,,
\label{order1-oscillator-expression-d}
\end{eqnarray}
and their Hermitian conjugates. 
We remind the reader that 
the zeroth components of the momentum in the plane-wave exponentials
in Eqs.~(\ref{order1-oscillator-expression-b}) and
(\ref{order1-oscillator-expression-d}) 
depend on the corresponding mode: 
$p^0_{u,s}$ and $p^0_{v,s}$.

The results derived in Appendix~\ref{canonical} 
for the free-field quantization 
in the presence of Lorentz violation 
hold analogously for the field $\psi_{in}(x)$.
Note in particular the (Feynman) propagator
\begin{widetext}
\begin{equation}
\langle0|T\psi_{in}(x)\bar\psi_{in}(y)|0\rangle=
\>\int\frac{d^4p}{(2\pi)^4}\frac{i\,e^{-ip\cdot(x-y)}}
{\Gamma^\mu p_\mu-m_{\rm ph}-m_c(c^{00}\omega_p^2
-2c^{0i}p^0p^i+c^{ij}p^ip^j)+i\epsilon}
\approx\int\frac{d^4p}{(2\pi)^4}\frac{i\,e^{-ip\cdot(x-y)}}
{\bar P(p)+i\epsilon}\,.
\label{in-propagator-linearized}
\end{equation}
\end{widetext}
In the last step, 
we have retained only the leading-order Lorentz-violating corrections
to the denominator of the integrand.
This approximation holds 
provided one ignores any possible unphysical poles 
far from the mass shell that 
might appear in the integration over $p^0$
when taking $\bar P(p)$ as the momentum-space two-point function.

From the discussion at the beginning of this section, 
we can now give a more precise formulation of the adiabatic hypothesis 
for the interacting field $\psi(x)$ and the free field $\psi_{in}(x)$.
Comparing Eq.~(\ref{in-propagator-linearized}) 
with the on-shell limit of Eq.~(\ref{GenPole}) 
it follows that
instead of relation~(\ref{LI-Z}) we now have,
in the limit $x_0\to-\infty$:
\begin{align}
\psi(x)\to&\int\!\frac{d^3\vec{p}}{(2\pi)^3}\,\frac{m}{\omega_p}
\,\mathcal{Z}^{{1 \over 2}}_R\bigl((c,\tilde k)^p_p\bigr)
\sum_{s=1}^2\big[
b^{in}_s(\vec p)u^{s}_{in,1}(\vec p)e^{-ip\cdot x}
\nonumber\\
&
\hspace{6em}+d_s^{in}{}^\dagger(\vec p)v^s_{in,1}(\vec p)e^{ip\cdot x}\big]\,.
\label{FmodeExpansion2}
\end{align}
For large negative times, 
Eq.~(\ref{order1-oscillator-expression-b})
can thus be written in terms of the interacting field,
\begin{equation}
b^{in}_s{}^\dagger(\vec p) = 
\int \!\! d^3x\,e^{-i p\cdot x}\bar\psi(x)\tilde\Gamma^0(\vec p\,)u^s_{in,1}(\vec p)
\mathcal{Z}^{-{1\over 2}}_R\bigl((c,\tilde k)^p_p\bigr)\,,
\label{order1-oscillator-expression-b2}
\end{equation}
and similarly for $d_{in}^\dagger$.
In the same way we can express 
the out-oscillators in terms of the interacting field 
for large positive times.

We will now use the above results 
to derive the LSZ reduction formula
for the Lorentz-violating case.
For this expression to exist, 
it is important that the Cauchy initial-value problem of the field theory 
be well defined.
Fortunately, 
the above procedure leading to a spinor equation of motion
linear in the zeroth component of the 4-momentum 
is exactly what is needed for a consistent derivation of the LSZ formula, 
as we will see below.

We begin by defining a smearing procedure 
for the definite-momentum creation and annihilation operators,
so that the states created by the smeared operators 
can be localized in position space. 
For example, 
\begin{equation}\label{creation-wave-packet}
b_{\vec q,s}^\dagger\equiv\int \widetilde{d^3 p}\,f_q(\vec p\,)b_s^\dagger(\vec p)\,,
\end{equation}
with analogous definitions for the various other creation and annihilation operators.
Here, 
we have abbreviated $\widetilde{d^3 p}\equiv (m \, d^3\vec{p})/(\omega_p 8\pi^3)$, 
and 
$f_q(\vec p\,)\sim\exp[-(\vec p-\vec q)^2/4\sigma^2]$,
describing the creation of a particle 
localized in 3-momentum space near $\vec q$
and localized in 3-position space near the origin.
In the Schr\"odinger picture, 
a state created by this operator evolves in time.
Applying this smearing to $b^{in}_s{}^\dagger$ 
given in the form of Eq.~(\ref{order1-oscillator-expression-b2})
yields
\begin{widetext}
\begin{align}
\label{evolution-bdagger1}
b^{in}_{\vec q,s}{}^\dagger&=\int\! \widetilde{d^3 p}\,f_q(\vec p\,)
\int \! d^3x\,e^{-i p\cdot x}\bar\psi(x)\tilde\Gamma^0(\vec p\,)u^s_{in,1}(\vec p)
\,\mathcal{Z}^{-{1\over 2}}_R\bigl((c,\tilde k)^p_p\bigr)
\vert_{t=-\infty}\nonumber\\
&=b^{out}_{\vec q,s}{}^\dagger-\int \! \widetilde{d^3 p}\,f_q(\vec p\,)\int\! d^4x\,
\partial_0\Big[e^{-i p\cdot x}\bar\psi(x)\tilde\Gamma^0(\vec p\,)u^s_{in,1}(\vec p)
\,\mathcal{Z}^{-{1\over 2}}_R\bigl((c,\tilde k)^p_p\bigr)\Big]\nonumber\\
&=b^{out}_{\vec q,s}{}^\dagger+i\!\int\! \widetilde{d^3 p}\,f_q(\vec p\,)\int\! d^4x\,
\bar\psi(x)\tilde\Gamma^0(\vec p\,)(i\hspace{-.25em}\lprt_0\hspace{-.33em}
{}+p^0)u^s_{in,1}(\vec p)\,e^{-ip\cdot x}
\mathcal{Z}^{-{1\over 2}}_R\bigl((c,\tilde k)^p_p\bigr)
\,.
\end{align}
We can now use the equation of motion~(\ref{eigenvalue-equation-Gamma0}) for $u^s_{in,1}(\vec p)$ 
to express $\tilde\Gamma^0p^0$ in the last equation in terms of $\vec p$.
We then trade the $p^i$ components for partial derivatives acting
to the right on the exponential.
By performing partial integrations 
they can be converted to partial derivatives acting to the left,
which yields
\begin{align}
&b^{in}_{\vec q,s}{}^\dagger-b^{out}_{\vec q,s}{}^\dagger=\nonumber\\
&=i\!\int\! \widetilde{d^3 p}\,f_q(\vec p\,)\int\! d^4x\,
\bar\psi(x)\biggl[i
\hspace{-.25em}\lprt_0\hspace{-.24em}
\tilde\Gamma^0
(i
\hspace{-.3em}\lnabla
)+i\Gamma^i
\hspace{-.25em}\lprt_i\hspace{-.33em}
{}+m_{\rm ph}
+m_c\left(c^{00}(m^2-
\hspace{-.3em}\lnabla{\hspace{-.25em}}^2\hspace{-0em}
)
-c^{ij}
\hspace{-.35em}\lprt_i \lprt_j
\right)\biggr]
u^s_{in,1}(\vec p)\,e^{-ip\cdot x}
\mathcal{Z}^{-{1 \over 2}}_R\bigl((c,\tilde k)^p_p\bigr)\nonumber\\
&=i\!\int\! \widetilde{d^3 p}\,f_q(\vec p\,)\int\! d^4x\,
\bar\psi(x)\biggl[i\tilde\Gamma^\mu
\hspace{-.3em}\lprt_\mu\hspace{-.4em}
{}+m_{\rm ph}
-m_c\left(c^{\mu\nu}
\hspace{-.35em}\lprt_\mu \lprt_\nu\hspace{-.4em}
{}-c^{00}(
\hspace{+.1em}\lsquare\hspace{-.2em}
{}+m^2)\right)\biggr]
u^s_{in,1}(\vec p)\,e^{-ip\cdot x}
\mathcal{Z}^{-{1 \over 2}}_R\bigl((c,\tilde k)^p_p\bigr)\nonumber\\
&\approx -i\!\int\! \widetilde{d^3 p}\,f_q(\vec p\,)\int\! d^4x\,
\bar\psi(x)\tilde P(-i
\hspace{-.25em}\lprt\hspace{.2em}
)
u^s_{in,1}(\vec p)\,e^{-ip\cdot x}
\mathcal{Z}^{-{1 \over 2}}_R\bigl((c,\tilde k)^p_p\bigr).
\label{evolution-bdagger3}
\end{align}
The last identity is valid to first order in Lorentz violation,
on the physical mass shell
(i.e., any spurious, unphysical solutions of $\bar P(p)=0$
far from the mass shell should be disregarded).
Similarly, 
if we start with an antifermion in the initial state:
\begin{align}
&d^{in}_{\vec q,s}{}^\dagger-d^{out}_{\vec q,s}{}^\dagger
\approx i\!\int\! \widetilde{d^3 p}\,f_q(\vec p\,)\int\! d^4x\,
\bar v^s_{in,1}(\vec p)\;e^{-ip\cdot x}
\tilde P(i\vec{\partial}\,)\psi(x)
\mathcal{Z}^{-{1 \over 2}}_R\bigl((c,\tilde k)^p_p\bigr)\,.
\label{evolution-ddagger}
\end{align}

Suppose we have fermions labeled ($p_1,\ldots$) and antifermions
labeled ($p'_1,\ldots$) in the in-state,
and fermions labeled ($q_1,\ldots$) and antifermions
labeled ($q'_1,\ldots$) in the out-state 
(label for spin degrees of freedom and vector arrows suppressed for clarity).
The conjugate spacetime variables are respectively denoted
$(x_1,\ldots)$, $(x'_1,\ldots)$, $(y_1,\ldots)$, and $(y'_1,\ldots)$.
It follows that the scattering amplitude,
\begin{equation}
\label{scattering1}
\langle f|i\rangle=
\langle\mathrm{out}|\cdots d^{out}_{q'_1}\cdots b^{out}_{q_1}
b^{in}_{p_1}{}^\dagger\cdots d^{in}_{p_1'}{}^\dagger\cdots|\mathrm{in}\rangle\,,
\end{equation}
can be expressed, 
using Eqs.~(\ref{evolution-bdagger3})
and~(\ref{evolution-ddagger}) and their Hermitian conjugates, 
as the LSZ reduction formula 
\begin{align}
\langle f|i\rangle=
&\int\! d^4x_1\cdots d^4y'\cdots 
\exp\big[-i(p\cdot x+\cdots+p'\cdot x'+\cdots
-q\cdot y-\cdots-q'\cdot y'-\cdots)\big]\nonumber\\
&\qquad{}\times
\cdots(-i)\mathcal{Z}^{-{1 \over 2}}_R\bigl((c,\tilde k)^{q_1}_{q_1}\bigr)
\bar u_{in}(\vec q_1)
\tilde P(i\vec{\partial}_{y_1})\cdots
i\mathcal{Z}^{-{1 \over 2}}_R\bigl((c,\tilde k)^{p_1'}_{p_1'}\bigr)
\bar v_{in}(\vec p\,'_{\!1})
\tilde P(i\vec{\partial}_{x'_1})\nonumber\\
&\qquad{}\times
\langle0|T[\cdots\bar\psi(y_1')\cdots\psi(y_1)
\bar\psi(x_1)\cdots\psi(x_1')\cdots]|0\rangle\nonumber\\
&\qquad{}\times
\tilde P(-i
\hspace{-.25em}\lprt_{\hspace{-.05em}x_1})
u_{in}(\vec p_1)
(-i)\mathcal{Z}^{-{1 \over 2}}_R\bigl((c,\tilde k)^{p_1}_{p_1}\bigr)\cdots
\tilde P(-i
\hspace{-.25em}\lprt_{\hspace{-.05em}y_1})
v_{in}(\vec q\,'_{\!1})
i\mathcal{Z}^{-{1 \over 2}}_R\bigl((c,\tilde k)^{q_1'}_{q_1'}\bigr)\cdots\nonumber\\
&{}+\mathrm{disconnected\ terms}.
\label{scattering2}
\end{align}
\end{widetext}
As in the derivation for the Lorentz-invariant case 
(see, e.g., Ref.~\cite{Itzykson}), 
the introduction of the time-ordered product in Eq.~(\ref{scattering2}) 
is necessary so that 
the field operators are in a convenient order 
with respect to the in- and out-vacua.
In deriving Eq.~(\ref{scattering2}), 
we have taken the momentum distributions
$f_q(\vec p\,)$ to the delta-function limit,
\begin{equation}\label{delta-limit}
f_q(\vec p\,)\to\delta^3(\vec p-\vec q).
\end{equation}

In practical calculations it is most useful to express the
scattering amplitude in terms of truncated Green's functions.
Using the definition
\begin{widetext}
\begin{equation}
\tilde G_{2n}(p_1',\ldots p_n';p_1,\ldots,p_n)=
\int\prod_{i=1}^n d^4z_i'\,d^4z_i\,
\exp\Big[i\sum_{i=1}^n (p_i'\cdot z_i'+p_i\cdot z_i)\Big]
T\Big[\prod_{i=1}^n\bar\psi(z_i')\prod_{j=1}^n\psi(z_j)\Big]
\label{Greens-function}
\end{equation}
in Eq.~(\ref{scattering2}),  
the connected scattering amplitude can be expressed as
\begin{align}
\langle f|i\rangle_c&=
\cdots(-i)\mathcal{Z}^{-{1 \over 2}}_R\bigl((c,\tilde k)^{q_1}_{q_1}\bigr)
\bar u_{in}(\vec q_1)
\bar P(-q_1)\cdots
i\mathcal{Z}^{-{1 \over 2}}_R\bigl((c,\tilde k)^{p_1'}_{p_1'}\bigr)
\bar v_{in}(\vec p\,'_{\!1})
\bar P(p'_1)
\tilde G^{(2n)}(-q'_1,\ldots,p_1,\ldots;-q_1,\ldots,p_1',\ldots)\nonumber\\
&\quad{}\times
\bar P(-p_1)
u_{in}(\vec p_1)
(-i)\mathcal{Z}^{-{1 \over 2}}_R\bigl((c,\tilde k)^{p_1}_{p_1}\bigr)\cdots
\bar P(q_1')
v_{in}(\vec q\,'_{\!1})
i\mathcal{Z}^{-{1 \over 2}}_R\bigl((c,\tilde k)^{q_1'}_{q_1'}\bigr)\cdots\,.
\label{scattering3}
\end{align}
If we now introduce the truncated Green's functions,
\begin{align}
(2\pi)^4&\delta\left(\sum p_i+p_i'\right)
G^{(2n)}_{\mathrm{\tiny trunc}}(p_1',\ldots p_n';p_1,\ldots,p_n)=
\prod_{i=1}^n \left[\Gamma^{(2)}(p_i')\,\Gamma^{(2)}(p_i)\right]
\tilde G_{2n}(p_1',\ldots p_n';p_1,\ldots,p_n)\,,
\label{truncated}
\end{align}
in which all external legs are multiplied by the inverses of the
corresponding complete propagators, 
it follows from Eqs.~(\ref{LVansatzPsymmExpan}),
(\ref{scattering3}), and~(\ref{Greens-function}) that
\begin{align}
\langle f|i\rangle_c&=(2\pi)^4\delta^4(\sum p_i+\sum p'_i
-\sum q_i-\sum q'_i)
\cdots(-i)\mathcal{Z}^{{1 \over 2}}_R\bigl((c,\tilde k)^{q_1}_{q_1}\bigr)
\bar u_{in}(\vec q_1)
\cdots i\mathcal{Z}^{{1 \over 2}}_R\bigl((c,\tilde k)^{p_1'}_{p_1'}\bigr)
\bar v_{in}(\vec p\,'_{\!1})
\nonumber\\
&\quad{}\times
G^{(2n)}_{\mathrm{\tiny trunc}}(-q'_1,\ldots,p_1,\ldots;-q_1,\ldots,p_1',\ldots)
u_{in}(\vec p_1)
(-i)\mathcal{Z}^{{1 \over 2}}_R\bigl((c,\tilde k)^{p_1}_{p_1}\bigr)\cdots
v_{in}(\vec q\,'_{\!1})
i\mathcal{Z}^{{1 \over 2}}_R\bigl((c,\tilde k)^{q_1'}_{q_1'}\bigr)\cdots \,.
\label{scattering4}
\end{align}
Note that $G^{(2n)}_{\mathrm{\tiny trunc}}$ carries $2n$ Dirac
indices (that are contracted with the spinors),
which are suppressed here for readability.
\end{widetext}

Formula~(\ref{scattering4}) embodies the Feynman rules for the
scattering amplitude, incorporating:
\begin{itemize}
\item a momentum-conserving delta-function;
\item the amputated Green's function;
\item a momentum-dependent wave-function renormalization
factor
$\pm i\mathcal{Z}^{{1 \over 2}}_R\bigl((c,\tilde k)^{p}_{p}\bigr)$
for every external leg;
\item a Dirac spinor for every external leg:
\begin{itemize}
\item $u^s_{in}(\vec p)$ for an incoming fermion;
\item $\bar u^s_{in}(\vec p)$ for an outgoing fermion;
\item $v^s_{in}(\vec p)$ for an outgoing antifermion;
\item $\bar v^s_{in}(\vec p)$ for an incoming antifermion.
\end{itemize}
\end{itemize}

We will end this section with a derivation of some explicit
formulas for the spinors $u^s_{in}(p)$ and $v^s_{in}(p)$
satisfying Eqs.~(\ref{modified-u-eom}) and~(\ref{modified-v-eom}).
The most convenient way to achieve this is to take them
proportional to the \textit{usual} Lorentz-invariant spinor functions
$u^s_{LI}(p)$ and $v^s_{LI}(p)$,
but then not calculated for the real, physical momentum $p^\mu$, but for
a redefined momentum value $\tilde p^\mu$ satisfying
\begin{equation}\label{tilde-p-condition}
\tilde{\slashed p}-m_{\rm ph}\propto\bar P(p)\,.
\end{equation}
Thus
\begin{eqnarray}\label{external-u}
u^s_{in}(p)&=&C_u(\vec p\,) u^s_{LI}(\tilde p)\\
v^s_{in}(p)&=&C_v(\vec p\,) v^s_{LI}(\tilde p)
\label{external-v}
\end{eqnarray}
where $C_u(\vec p\,)$ and $C_v(\vec p\,)$ are normalization constants 
to be determined below.
One easily checks that
\begin{align}
\tilde p^\mu &=
\left(1-\frac{m_c}{m_{\rm ph}}c^p_p-\frac{m_k}{m_{\rm ph}}\tilde k^p_p\right)p^\mu
+\left(\bar x  c^{\mu\nu}+\bar y  \tilde k^{\mu\nu}\right)p_\nu\nonumber\\ 
&=\left(1+\frac{\alpha}{3\pi m^2}(2c^p_p-\tilde k^p_p)\right)p^\mu+(c_\phy)^{\mu\nu}p_\nu
\label{tilde-p-explicit}
\end{align}
satisfies Eq.~(\ref{tilde-p-condition})
to first order in the Lorentz-violating parameters 
and obeys the dispersion relation
$\tilde p^2=m_{\rm ph}^2$.

Let us work out the normalization
constants $C_u(\vec p\,)$ and $C_v(\vec p\,)$, in accordance with
Eq.~(\ref{order1-norm-u,v}).
Consider the case of $C_u(\vec p\,)$ first.
For $u^s_{LI}(\vec p)$ we have the usual relations
\begin{eqnarray}\label{LI-normalization1}
\bar u^r_{LI}(\tilde p)\gamma^\mu u^s_{LI}(\tilde p)
&=&\frac{\tilde p^\mu}{m_{\rm ph}} \delta_{rs}\,,\\
\bar u^r_{LI}(\tilde p) u^s_{LI}(\tilde p)
&=&\delta_{rs}\,.
\label{LI-normalization2}
\end{eqnarray}
Demanding now that $u^s_{in}(p)$ satisfies 
the normalization condition~(\ref{order1-norm-u,v}) 
it follows that
\begin{equation}
|C_u(\vec p\,)|^2\bar u^r_{LI}(\tilde p)\tilde\Gamma^0(\vec p\,)
u^s_{LI}(\tilde p)=\frac{\omega_p}{m}\delta_{rs}\,.
\end{equation}
Using Eqs.~(\ref{LI-normalization1}) and~(\ref{LI-normalization2})
and working to first order in Lorentz violation
one obtains the following expression for the normalization constant:
\begin{align}
&|C_u(\vec p\,)|^{-2}\nonumber\\
={}&\frac{m}{\omega_p\, m_{\rm ph}}\left[\tilde p^0+(c_\phy)^{0\nu}\tilde p_\nu
+\frac{2\alpha}{3\pi}(2c^{0i}-\tilde k^{0i})\tilde p^i\right]\nonumber\\
={}&\frac{m\,\omega_{\tilde p}}{\omega_p\, m_{\rm ph}}\left[1+\frac{1}{\omega_{\tilde p}}
\left((c_\phy)^{0\nu}\tilde p_\nu
+\frac{2\alpha}{3\pi}(2c^{0i}-\tilde k^{0i})\tilde p^i\right)\right]\,.
\label{norm-C_u}
\end{align}
In the last equation, 
we defined
$\omega_{\tilde p}\equiv\tilde p^0=\sqrt{\tilde p^i\tilde p^i+m_{\rm ph}^2}$.
Note that 
the same analysis can be done for the $v^s_{in}(p)$ spinors.
The normalization constant turns out to be the same as for the $u$ spinors,
so that we can safely suppress the $u$ and $v$ indices:
\begin{equation}
C_v(\vec p\,)=C_u(\vec p\,)\equiv C(\vec p\,)\,.
\end{equation}
As an additional simplification,
the normalization constants are also independent of the spin index $s$.
This allows us to determine spin-sum formulas.
They follow directly from the usual expressions for the Lorentz-invariant
case:
\begin{eqnarray}
\label{spin-sum-u}
\sum_{s} u^s_{in}(\vec p)\bar u^s_{in}(\vec p)&=&|C(\vec p\,)|^2\,
\frac{\tilde{\slashed p}+m_{\rm ph}}{2m_{\rm ph}}\\
\sum_{s} v^s_{in}(\vec p)\bar v^s_{in}(\vec p)&=&|C(\vec p\,)|^2\,
\frac{\tilde{\slashed p}-m_{\rm ph}}{2m_{\rm ph}}
\label{spin-sum-v}
\end{eqnarray}
In Eqs.~(\ref{spin-sum-u}) and~(\ref{spin-sum-v}), 
it is understood that $p^0\equiv p^0_u=p^0_v$
[see Eqs.~(\ref{eom_u_quadratic}) and~(\ref{eom_v_quadratic})].


\section{Sample calculation: infrared divergences in Coulomb scattering}
\label{Coulomb-scattering}

It is instructive to apply the techniques described above 
to a particular case.
We will do this for the Coulomb (or rather Mott) scattering
of a fermion off a stationary charge.
For simplicity, 
we will assume that 
only the Lorentz-violating parameter $\tilde k^{\mu\nu}$ is nonzero.

Let us review quickly the Lorentz-invariant case.
We have for the scattering amplitude at tree level
\begin{equation}
S_{fi}=ie
\frac{m}{V\sqrt{E_i E_f}}\int\! d^4x\, \bar u^s(p_f)
\slashed A(x)e^{i(p_f-p_i)\cdot x}u^r(p_i)\,.
\end{equation}
Here, 
we have normalized the states in a finite volume $V$.
For the Coulomb problem,
we can take
$\vec A=0$ and $A_0=Ze/4\pi|\vec x|$, so
\begin{align}
S_{fi}=&\frac{iZ\alpha}V \frac{m}{\sqrt{E_i E_f}}2\pi\delta(E_f-E_i)
\nonumber\\
&\qquad{}\times\int \! d^3x\, \frac{e^{-i\vec q\cdot \vec x}}{|\vec x|}\,
\bar u^s(p_f)\gamma^0 u^r(p_i)\,.
\label{S_fi}
\end{align}
We can now pass to the cross section by squaring the
absolute value of Eq.~(\ref{S_fi}),
multiplying by the number of possible final states $V\,d^3p_f/(2\pi)^3$
and dividing by the incident flux $|\vec v_i|/V$
and the time interval $T$.
Note that for large time intervals $T$, one can take 
$|2\pi\delta(E_f-E_i)|^2 \equiv T2\pi\delta(E_f-E_i)$.
It then follows that
\begin{align}
d\sigma_{fi}^0&=\int\frac{4Z^2\alpha^2m^2}{|\vec p_i|E_f|\vec q\,|^4}\,
\delta(E_f-E_i)\nonumber\\
&\quad{}\times|\bar u^s(p_f)\gamma^0 u^r(p_i)|^2
p_f^2dp_f d\Omega_f\,.
\label{LI-sigma_fi-1}
\end{align}
Using now that
\begin{equation}
|\vec p_i|=|\vec p_f|=p_f\qquad\mbox{and}\qquad
p_fdp_f=E_fdE_f\,,
\label{LI-E-p}
\end{equation}
it follows that
\begin{equation}
d\sigma_{fi}^0=\frac{4Z^2\alpha^2m^2}{|\vec q\,|^4}
|\bar u^s(p_f)\gamma^0 u^r(p_i)|^2 d\Omega_f\,.
\label{LI-sigma_fi-2}
\end{equation}
If we do not observe the final polarization,
we must sum over $s$,
while for an unpolarized incident wave we average over the
initial polarizations $r$.
With the usual formulas for the spin sums one obtains
\begin{align}
\left.\frac{d\sigma_{fi}^0}{d\Omega}\right\vert_{\mbox{\tiny unpol}}&=
\frac{4Z^2\alpha^2m^2}{|\vec q\,|^4}
\frac12\mbox{tr}\left(\gamma^0\frac{\slashed p_i+m}{2m}
\gamma^0\frac{\slashed p_f+m}{2m}\right)\nonumber\\
&=\frac{Z^2\alpha^2}{4|\vec p\,|^2\beta^2\sin^4(\theta/2)}
\left(1-\beta^2\sin^2\frac{\theta}{2}\right)\,.
\label{LI-sigma_fi-unpol}
\end{align}

When turning on Lorentz violation, various adaptations have
to be made to the formulas~(\ref{S_fi})--(\ref{LI-sigma_fi-unpol})
at tree level:
\begin{enumerate}
\item The Maxwell equations become Lorentz violating:
\begin{equation}
\square A^\mu=j^\mu\quad\Rightarrow\quad\tilde\square\,
\tilde\eta^{\mu\nu}A_\nu=j^\mu\,,
\label{Maxwell}
\end{equation}
where $\tilde \eta^{\mu\nu}= \eta^{\mu\nu}+\tilde k^{\mu\nu}$
and $\tilde\square=\partial_\alpha\tilde\eta^{\alpha\beta}\partial_\beta$.
This means that the Fourier transform of the Coulomb potential
becomes
\begin{equation}
{}\hspace{2em}\hat A_\mu=Ze\frac{\delta^0_\mu-\tilde k^0_\mu}
{q_\alpha\tilde\eta^{\alpha\beta}q_\beta}\delta(q^0)=
Ze\frac{\delta^0_\mu-\tilde k^0_\mu}
{q_i\tilde \eta^{ij} q_j}\delta(q^0)\,.
\end{equation}

\item The incident velocity is now given by the group velocity
$v_i^g=\partial E/\partial p^i$, which is fixed by the
dispersion relation~(\ref{dispersion-relation}).
However, note that,  as  in this example 
we choose $c^{\mu\nu}=0$,  there is no Lorentz-violating
effect at tree level.

\item The dispersion relation~(\ref{dispersion-relation}) also
implies Lorentz-violating modifications to the integration-variable transformation from
$dp_f$ to $dE_f$ implied by Eq.~(\ref{LI-E-p}).
However, also here there is no effect at tree level
because we take $c^{\mu\nu}=0$.
Incidentally, 
note that the factors $\sqrt{E_i}$ and $\sqrt{E_f}$ in the
denominator of Eq.~(\ref{S_fi}) remain equal to their Lorentz-invariant
form $\sqrt{\omega_{i,f}}$ ($\omega\equiv\sqrt{\vec p\,^2+m^2}$).

\item The spinors are modified according to the relations
(\ref{external-u})--(\ref{tilde-p-explicit}) and
(\ref{norm-C_u}).
In the unpolarized cross section~(\ref{LI-sigma_fi-unpol}),
we have to use the modified spin sum~(\ref{spin-sum-u})
or~(\ref{spin-sum-v}), as appropriate.
Again, there is no effect at tree level as we take $c^{\mu\nu}=0$.
\end{enumerate}

\begin{center}
\begin{figure*}
\includegraphics[width=0.70\hsize]{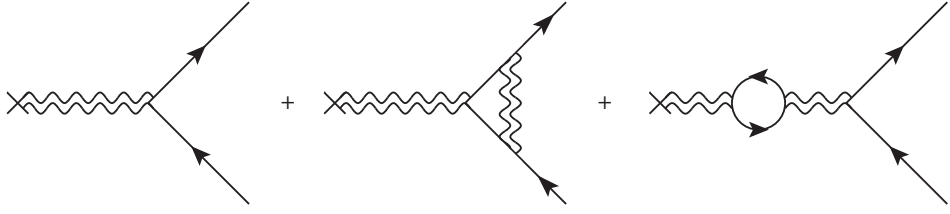}
\caption{Diagrammatic topologies contributing to the vertex function up to 
 one-loop order with $c^{\mu \nu} = 0$. Counterterm contributions have been omitted for clarity.}
\label{scatfig}
\end{figure*}
\end{center}

\vskip-20pt
It is straightforward but tedious to adapt the formulas
for the tree-level cross sections (\ref{LI-sigma_fi-2})
and (\ref{LI-sigma_fi-unpol}) to the
Lorentz-violating case accordingly.
Rather than doing this explicitly,
we will move our attention to radiative corrections.
The diagrams shown in Fig.~\ref{scatfig} contribute to one-loop order.
Note that the fermion self-energy diagrams are taken into account
implicitly in the order $\alpha$ corrections to the
external spinors in formulas~(\ref{external-u})--(\ref{norm-C_u}),
as well as by the inclusion of the wave-function
renormalization $\mathcal{Z}_R^{1/2}$ for each external
fermion leg.

Instead of carrying out the full calculation of the one-loop diagrams,
we will just concentrate on the infrared-divergent contributions
to the scattering amplitude.
We will then show that 
they indeed cancel in the experimental cross section,
just as in the Lorentz-invariant case.

Paralleling the usual Lorentz-invariant case, 
the vacuum-polarization diagram is infrared finite, 
so that we only have to consider the
vertex-correction diagram.
With the modified photon propagator 
(recall that we have chosen $c^{\mu\nu}=0$)
the amplitude for the vertex correction is given by
\begin{widetext}
\begin{align}
-iq\Gamma^\mu(p',p)=(-iq)^3\mu^\epsilon\int\frac{d^dk}{(2\pi)^d} 
\frac{\gamma^\alpha i(\slashed p'-\slashed k+m)\gamma^\mu 
i(\slashed p-\slashed k+m)\gamma^\beta
(-i)(\eta_{\alpha\beta}-\tilde k_{\alpha\beta})}
{\bigl((p'-k)^2-m^2+i\epsilon\bigr)\bigl((p-k)^2-m^2+i\epsilon\bigr)
\bigl(k^2+2\tilde k^k_k-m_\gamma^2+i\epsilon\bigr)}\,.
\label{vertex-correction1}
\end{align}
The integral on the right-hand side of Eq.~(\ref{vertex-correction1}) is infrared divergent.
This divergence arises from the residue of the photon propagator.
In order to isolate it,
we can use the fermion on-shell relations
$p^2-m^2=\mathcal O(\alpha)\approx 0$ and
$(p')^2-m^2=\mathcal O(\alpha)\approx 0$.
It is then straightforward to show that
\begin{align}
\bar u^r(p') \gamma^\alpha(\slashed p'+m)\gamma^\mu(\slashed p+m)
\gamma^\beta 
u^s(p) (\eta_{\alpha\beta}-\tilde k_{\alpha\beta})
=4 \bar u^r(p')
\gamma^\mu 
u^s(p) 
(p'\cdot p-\tilde k^{p'}_p)
\end{align}
by using the (approximate) equations of motion for the spinors.
Moreover, 
we can drop terms linear in $k$ in the numerator
and quadratic in $k$ in the fermion pole factors in the denominator.
It then follows that
\begin{align}
\bar u^r(p')
\Gamma^\mu(p',p)
u^s(p)
\approx
-iq^2\mu^\epsilon
\bar u^r(p')
\gamma^\mu
u^s(p)
\int\frac{d^dk}{(2\pi)^d}\frac{p'\cdot p-\tilde k^{p'}_p}
{k^2+2\tilde k^k_k-m_\gamma^2+i\epsilon}
\frac{1}{(p'\cdot k)(p\cdot k)}\,.
\label{vertex-correction2}
\end{align}
We can evaluate the integral~(\ref{vertex-correction2}) by using the identity
\begin{equation}
\frac{1}{k^2+2\tilde k^k_k-m_\gamma^2+i\epsilon}=
\mbox{P}\frac{1}{k^2+2\tilde k^k_k-m_\gamma^2}
-i\pi\delta(k^2+2\tilde k^k_k-m_\gamma^2)\,,
\label{identity-pole-PP}
\end{equation}
where P denotes the principle-value part. 
Omitting the  infrared-finite contribution from the principal value
we find
\begin{align}
\bar u^r(p')
\Gamma^\mu(p',p)
u^s(p)
=
-\frac{\alpha}{4\pi^2}
\bar u^r(p')
\gamma^\mu
u^s(p)
\int d^4k\,\delta(k^2+2\tilde k^k_k-m_\gamma^2)\frac{p'\cdot p-\tilde k^{p'}_p}
{(p'\cdot k)(p\cdot k)}.
\end{align}

It follows that the one-loop contributions to the
elastic Coulomb scattering amplitude are obtained by substituting
$\mathcal{M}_0\equiv
\bar u^{s_f}(p_f)
\gamma^0
u^{s_i}(p_i)$
in Eq.~(\ref{LI-sigma_fi-2}) by
\begin{equation}
\mathcal{M}=\mathcal{M}_0\left\{1-\frac{\alpha}{4\pi^2}\int d^4k
\delta(k^2+2\tilde k^k_k-m_\gamma^2)
\frac{p'\cdot p-\tilde k^{p'}_p}{(p'\cdot k)(p\cdot k)}\right\}
\mathcal{Z}_R^{1/2}(p')\mathcal{Z}_R^{1/2}(p)+\cdots\,.
\end{equation}
where the ellipsis indicates infrared-finite contributions.
For the elastic cross section, 
this implies
\begin{equation}
\frac{d\sigma_{fi}^{el}}{d\Omega}=\frac{d\sigma_{fi}^0}{d\Omega}
\left\{1-\frac{\alpha}{2\pi^2}\int d^4k
\delta(k^2+2\tilde k^k_k-m_\gamma^2)
\frac{p'\cdot p-\tilde k^{p'}_p}{(p'\cdot k)(p\cdot k)}\right\}
\mathcal{Z}_R(p')\mathcal{Z}_R(p)+\cdots\,.
\label{sigma_fi-el}
\end{equation}
\end{widetext}
The term proportional to $\alpha$ as well as the factors $\mathcal{Z}_R$
in expression~(\ref{sigma_fi-el}) are infrared divergent for $m_\gamma\to0$.

In the Lorentz-invariant case, 
infrared divergences are canceled
if one incorporates the fact that some final states that include soft photons
are experimentally indistinguishable.
We will now proceed to check that 
this continues to hold true in the case at hand.

Following the usual procedure,
we consider final states that 
include one soft photon
with energy smaller than 
the detector resolution $\Delta E$.
The amplitude for this process is
\begin{equation}
\mathcal{M}=q\mathcal{M}_0\left[\frac{2p'\cdot\epsilon_r}{2p'\cdot k}
+\frac{2p\cdot\epsilon_r}{-2p\cdot k}\right]\,,
\label{1gamma-amplitude}
\end{equation}
where $\mathcal{M}_0$ is the amplitude for elastic scattering
without photon emission.
To get the cross section for this process, we have to include
also the final-state volume element for the photon.
As shown in Appendix \ref{canonical-radiation},
a consistent way to do this is to take
\begin{equation}
\frac{d^3k}{(k_{0+}+k_{0-})(2\pi)^3}
\label{photon-final-state-element}
\end{equation}
rather than the conventional factor $d^3k/(2\omega_k(2\pi)^3)$.
Here, 
$k_{0\pm}(\vec k)$ are the absolute values of the solutions
for $k_0$ to the modified dispersion relation.

Combining Eqs.~(\ref{1gamma-amplitude}) and (\ref{photon-final-state-element})
and employing formula~(\ref{measure-property}) 
one then finds
for the cross section for this (soft-bremsstrahlung) process
\begin{widetext}
\begin{equation}
\frac{d\sigma_{fi}^{1\gamma}}{d\Omega}\approx
\frac{d\sigma_{fi}^{0}}{d\Omega}
\frac{q^2}{2(2\pi)^3}\int_{\omega\le\Delta E}d^4k\,
\tilde\eta^{00}\delta^4\bigl(k^2+\tilde k^k_k-m_\gamma^2\bigr)
\sum_{\lambda=1}^3
\left[\frac{2p'\cdot\epsilon^{(\lambda)}(k)}{2p'\cdot k}
+\frac{2p\cdot\epsilon^{(\lambda)}(k)}{-2p\cdot k}\right]^2.
\label{sigma_fi-1gamma-1}
\end{equation}
Using the polarization-sum formula~(\ref{polarization-sum2}),
Eq.~(\ref{sigma_fi-1gamma-1}) reduces to
\begin{equation}
\frac{d\sigma_{fi}^{1\gamma}}{d\Omega}\approx
\frac{d\sigma_{fi}^{0}}{d\Omega}
\frac{\alpha}{(2\pi)^2}\int_{\omega\le\Delta E}d^4k\,
\delta^4\bigl(k^2+\tilde k^k_k-m_\gamma^2\bigr)
\left[-\frac{(p')^2-\tilde k^{p'}_{p'}}{(p'\cdot k)^2}
-\frac{p^2-\tilde k^p_p}{(p\cdot k)^2}
+\frac{2p'\cdot p-2\tilde k^{p'}_p}{(p'\cdot k)(p\cdot k)}\right].
\label{sigma_fi-1gamma-2}
\end{equation}
\vskip1em
\end{widetext}
Comparing the third term inside brackets in Eq.~(\ref{sigma_fi-1gamma-2})
with Eq.~(\ref{sigma_fi-el})
we see that both contain integrals over the same $p,p'$ term,
with opposite sign, so that the corresponding infrared divergences
cancel.

To verify that the same happens for the $p,p$ and the $p',p'$ terms
in Eq.~(\ref{sigma_fi-1gamma-2}) and the
$\mathcal{Z}_R(p')\mathcal{Z}_R(p)$ factors in
Eq.~(\ref{sigma_fi-el}),
we evaluate the $k$ integral over, say, the  $p,p$ term
in Eq.~(\ref{sigma_fi-1gamma-2}).
To this end, 
we perform
a change of variables $k^\mu\to\bar k^\mu$ such that
\begin{equation}
\bar k^2=k^2+\tilde k^k_k\,.
\end{equation}
To first order, 
this means that
$\bar k^\mu=k^\mu+\frac12 \tilde k^{\mu\nu}k_\nu$.
To lowest order,
the measure $d^4k$ is invariant under this transformation, as
\begin{equation}
\mbox{Det}\left[\frac{\partial\bar k}{\partial k}\right]
\approx \mbox{Det}\Bigl[\delta^\mu_\nu+\frac12\tilde k^\mu_\nu\Bigr]
\approx1+\frac12\tilde k^\mu_\mu=1\,.
\end{equation}
It follows that
\begin{align}
\int_{\omega\le\Delta E}d^4k
\,\delta\bigl(k^2+\tilde k^k_k-m_\gamma^2\bigr)
\frac{p^2-\tilde k^p_p}{(p\cdot k)^2}
\nonumber\\
\approx\int_{\bar\omega\le\Delta E} d^4\bar k\,\delta(\bar k^2-m_\gamma^2)
\frac{\tilde p^2}{(\bar k\cdot\tilde p)^2}\,,
\label{integration-transform}
\end{align}
where we have defined $\tilde p^\mu=(\eta^{\mu\nu}-\frac12\tilde k^{\mu\nu})p_\nu$.
Note that strictly speaking, 
the upper limit of the transformed energy $\bar\omega$
is modified by the transformation, 
but the result is an effect of higher order
in $\tilde k$ and can be ignored.
The resulting integration is a standard one with result,
\cite{Itzykson}
\begin{equation}
4\pi\ln\left(\frac{\Delta E}{m_\gamma}\right)+\cdots\,,
\end{equation}
where the ellipsis indicates terms that 
are finite as $m_\gamma \to 0$.
Substitution in Eq.~(\ref{sigma_fi-1gamma-2}) gives
(for the second term)
\begin{equation}
-\frac{d\sigma_{fi}^0}{d\Omega}\,\frac{\alpha}{\pi}
\ln\left(\frac{\Delta E}{m_\gamma}\right)
+\mbox{finite}.
\label{sigma_fi-1gamma-2-second-term}
\end{equation}
It follows that the infrared divergence in
Eq.~(\ref{sigma_fi-1gamma-2-second-term}) indeed
cancels the corresponding one in the elastic cross section (\ref{sigma_fi-el})
arising from the multiplicative renormalization function
$\mathcal{Z}_R(p)$ that 
was evaluated in Eq.~(\ref{Z_R-2}):
\begin{equation}
\mathcal{Z}_R\bigl((c,k)^p_p\bigr)=1+\frac{\alpha}{\pi}
\ln\left(\frac{m}{m_\gamma}\right)+\mbox{finite}.
\end{equation}


\section{Summary and outlook}
\label{conclusions}

Perturbative Lorentz-invariant quantum field theory 
rests on a few core field-theoretic techniques.
One of these concerns the order-by-order determination 
of the asymptotic Hilbert space, 
and thus the calculation of quantum corrections 
to the external states.
Such effects govern the propagation of free particles 
and are indispensable for scattering amplitudes.
The present work for the first time 
has addressed the issue
how to generalize this core technique 
to Lorentz-violating quantum-field theories.
To illustrate the salient features of this generalization,
we have focused on a particular sector of the SME.
We expect, 
though,
that our reasoning can also be applied to other Lorentz-violating field theories
with a more complex structure and a wider variety of coefficients.

Specifically,
we considered the SME's single-flavor QED sector 
in the presence of the $c^{\mu\nu}$ and the nonbirefringent piece
of the $k_F^{\mu\nu\alpha\beta}$ coefficients.
Working perturbatively, 
we found that 
the presence of these Lorentz-breaking terms in the Lagrangian 
has some profound consequences for the radiative corrections
to the pole structure of the external states.
In particular,
the Dirac equation satisfied by the external-state spinors 
turns out to be modified by
Lorentz-violating operators not present in the Lagrangian,
a feature that is unknown in usual Lorentz-symmetric field theories.
Our analysis also shows that
the wave-function renormalization 
will typically 
contain Lorentz-breaking coefficients 
contracted with momenta. 
We note that this is in contrast to the usual one-loop QED result,
where $Z_\psi$ is a momentum-independent constant.
Momentum dependence of the wave-function renormalization
is known to occur in certain other contexts
\cite{momemtum_dep_Z}.

We have limited our present study primarily to theoretical techniques 
for determining quantum corrections 
to external states in Lorentz-violating backgrounds. 
However, 
our results indicate that 
such corrections may have profound phenomenological implications 
for Lorentz tests, 
which can be seen as follows. 
The new, radiatively induced term exhibits two powers of the momentum,
whereas the existing terms contain only up to a single power. 
The correction term should therefore grow faster with the momentum than 
the existing terms. 
This opens the possibility---at least in principle---that 
the Lorentz-breaking radiative corrections become larger than 
the original tree-level Lorentz violation. 
Note that 
this does not necessarily signal a breakdown of perturbation theory 
because the perturbation Hamiltonian 
also includes the conventional electromagnetic interaction, 
which is comparatively much larger.

In our simple model, 
the radiative-correction term of size $\sim\frac{\alpha}{3\pi m}(2c^p_p-\tilde{k}^p_p)$
can reach the size of the tree-level contribution $c^p_\gamma$
when $\alpha\, p\sim m$.
It follows that
for Lorentz tests involving free electrons with energies $\gsim 100\;$MeV,
radiative Lorentz-breaking corrections 
may not always be negligible 
relative to the tree-level Lorentz violation. 
Note that 
electrons in such an energy range
are routinely employed in various Lorentz tests.
We remark,
however, 
that in our particular model
the resulting fermion eigenenergies 
are free of this effect. 
This may be a special property of our model
as both the tree-level Lorentz violation 
and the induced correction 
have the same C, P, and T properties. 
Nevertheless, 
the model discussed in this work 
could still exhibit other observables,
such as ones involving the one-loop eigenspinors, 
in which the Lorentz-breaking correction 
dominates the tree-level effects. 
In the context of more general models 
involving the parity-odd weak interaction,
the radiatively induced terms 
are likely to display a greater variety of structures 
since they do not necessarily have to share 
the same C, P, and T properties of the tree-level 
Lorentz violation.
Then, 
even the eigenenergies 
may show the effects mentioned above.

Another immediate consequence of our result
concerns multimetric theories~\cite{multimetric}, 
such as recently proposed bimetric models~\cite{bimetric,bimcrit}.
The basic idea in models of this type is that
different fields experience different effective metrics. 
But our analysis shows that
the concept of two metrics is difficult to maintain in a quantum theory:
beyond tree level, 
radiative corrections to particle propagation 
typically induce higher-order terms 
incompatible with an effective-metric interpretation. 
This difficulty by itself does not affect the consistency of such models;
it rather illustrates, for example, that 
the trajectory of the particle 
is {\em not} a geodesic with respect to some metric.

To see this more explicitly, 
consider first 
the free electromagnetic field in our model. 
Inspection of Eqs.~(\ref{delta_eta}) and~(\ref{bare-model-lagrangian})
establishes that 
$\hat{\eta}^{\mu\nu}=\eta^{\mu\nu}+\tilde{k}^{\mu\nu}$
can be interpreted as the effective (inverse) metric 
that governs photon propagation at tree level.
Similarly,
comparison of our fermion kinetic term 
$\half i \overline{\psi}(\delta^{\mu}_a+c^{\mu}_a)\gamma^a\!\!\!\stackrel{\;\leftrightarrow}{\partial}_{\!\!\mu}\!\!\psi$ 
with that in general coordinates 
$\half i e \overline{\psi}e^{\mu}_a\gamma^a\!\!\!\stackrel{\;\leftrightarrow}{\partial}_{\!\!\mu}\!\!\psi$ 
reveals that 
we may interpret $\delta^{\mu}_a+c^{\mu}_a$ 
as the vierbein $e^{\mu}_a$~\cite{c=const_comment}.
It is then apparent that 
the fermion propagation
is controlled by the (inverse) effective metric 
$(g_f)^{\mu\nu}=e^\mu_a e^\nu_b \eta^{ab}=\eta^{\mu\nu}+2c^{\mu\nu}+{\cal O}(c^2)$ 
at tree level.
We see that
in the absence of quantum corrections 
our Lorentz-violating QED extension 
can indeed be interpreted as 
a bimetric model in the flat-spacetime limit.
We remark in passing that 
this is consistent with our earlier discussion 
that only $2c^{\mu\nu}-\tilde{k}^{\mu\nu}$
is observable.
On the other hand,
our analysis has shown that 
the leading radiative corrections to the free-fermion propagation---displayed
in Eq.~(\ref{P-phys})---are determined by a term of the form
$\overline{\psi}(2c^{\mu\nu}-\tilde{k}^{\mu\nu})\partial_\mu\partial_\nu\psi$. 
But such a term 
precludes an interpretation of the fermion's propagation 
as being governed by an effective metric.

On a more practical level, 
we applied our formalism to Coulomb scattering for the case
$c^{\mu\nu}=0$,
$\tilde{k}^{\mu\nu}\ne 0$.
We showed that, 
just like in the usual Lorentz-symmetric case,
infrared divergences cancel when soft-photon emission
is taken into account in the final fermion states.
It should be stressed that 
this result involves a nontrivial cancellation between
various infrared-divergent Lorentz-violating terms.
Our study also demonstrates 
how to extract the S-matrix of a process with external fermion states
in the presence of Lorentz-breaking coefficients,
generalizing the usual LSZ reduction formula 
to
incorporate leading-order SME effects.

In this paper, 
we have only considered
{\it linear}
Lorentz-violating radiative effects on {\it fermion} external states,  
which suggests two natural avenues for further exploration.
One of these concerns
a nonperturbative treatment of asymptotic states 
at all orders in Lorentz breaking 
within formal field theory. 
A study of this kind 
could yield additional theoretical insight, 
such as the analytic structure of the single-particle propagator 
away from the pole.
The second avenue for future investigations
involves photon-propagation effects,
and in particular the incorporation of Lorentz breakdown 
into vacuum polarization.
An analysis along these lines
needs to include a study of the K\"all\'en--Lehmann representation 
for the photon propagator
in the presence of the full $k_F^{\mu\nu\alpha\beta}$ coefficient
to facilitate a proper extraction of the photon pole(s).
We expect to return to this issue in the future.


\acknowledgments 
We thank Gregory Adkins, Alan Kosteleck\'y, Jacob Noord\-mans,
and Keri Vos for discussions.
R.P.~and M.C.\ acknowledge the kind hospitality 
of the Indiana University Physics Department,
and R.L.~the warm hospitality of Universidad Andr\'es Bello
during several stages of this study.
This work has been supported in part 
by the Portuguese Funda\c c\~ao para a Ci\^encia e a Tecnologia, 
by the Mexican RedFAE,
by Universidad Andr\'es Bello under Grant No.\ DI-27-11/R,
by FONDECYT under Grant No.\ 11121633,
by the Indiana University Center for Spacetime Symmetries,
by the IU Collaborative Research and Creative Activity Fund of the Office of the Vice President for Research, 
and by the IU Collaborative Research Grants program.

\appendix

\section{Feynman rules for the second perturbation scheme}
\label{2pointrules}

\begin{center}
\begin{figure}
\includegraphics[width=0.70\hsize]{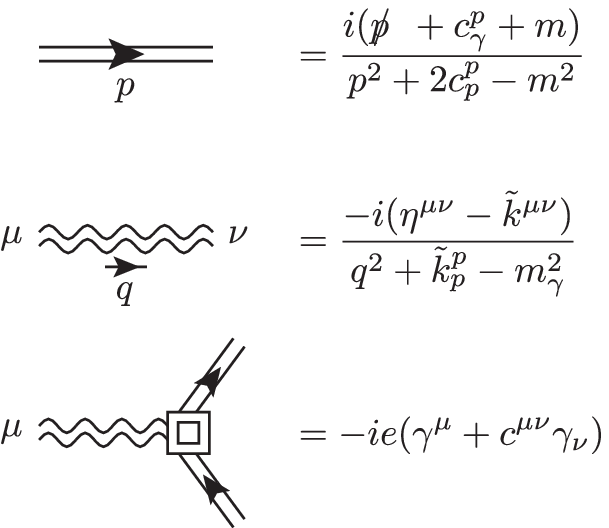}
\caption{Feynman rules in $\xi = 1$ gauge 
for the decomposition of our Lagrangian 
according to Eqs.~(\ref{L0}), (\ref{L1}), and  (\ref{L2}). 
Counterterm insertions are not shown.}
\label{FRules}
\end{figure}
\end{center}

This appendix presents the Feynman rules 
needed for perturbative calculations in our model. 
In Sec.~\ref{2pointfunction}, 
we briefly discussed two natural schemes 
for setting up perturbation theory,
each entailing different decompositions of  our model's full Lagrangian 
and the ensuing different sets of Feynman rules for each scheme.  
For convenience, 
we selected as the zeroth-order system 
the full renormalized quadratic Lagrangian 
(including quadratic Lorentz-violating pieces) 
and to treat the nonquadratic terms as a perturbation. 
The Feynman rules for this choice, 
i.e., 
for decomposition corresponding to Eqs.~(\ref{L0}), (\ref{L1}), and  (\ref{L2}), are displayed in Fig.~\ref{FRules},
where we have selected $\xi = 1$ Feynman gauge.
Counterterm expressions,
which are not displayed here,
have been taken from Ref.~\cite{Kostelecky:2001jc}.

\section{Expansion of $\bm{\Gamma^{(2)}(\bar P)}$}
\label{orderingApp}

We begin by considering the expression~(\ref{LVansatzP}) for 
$\Gamma^{(2)}(\bar P)$, 
which shows that expansions of the scalar coefficient functions
$A$, $C$, $K$, and $M$ are needed,
\begin{align}\label{LV_A_expansion}
A(\bar P^2+&2\bar m \bar P+\bar \beta,c^p_p,\tilde{k}^p_p)=\nonumber\\
&
\sum_{n=0}^\infty\frac{1}{n!}A^{(n)}(\bar \beta,c^p_p,\tilde{k}^p_p)(\bar P^2+2\bar m \bar P)^n\,,
\end{align}
with analogous expressions for $C$, $K$, and $M$. 
Here, 
we have denoted the $n$th derivative with respect to the first argument 
by the superscript $(n)$.
No ordering ambiguities arise at this stage,
because the only nontrivial matrix in all of these expansions is $\bar P$.

More care is required 
when the expansions for $A$, $C$, $K$, and $M$ 
are inserted into 
the expression~(\ref{LVansatzP}) for $\Gamma^{(2)}(\bar P)$ 
because the expansions are then multiplied by 
$c^p_\gamma$ and $\tilde{k}^p_\gamma$, 
and these two matrices do in general not commute with $\bar P$.
However, 
by virtue of Eq.~(\ref{p-squared-bar-P}) 
we see that $(\bar P^2+2\bar m \bar P)=p^2-\bar\beta$ 
is proportional to $\openone$ and thus commutes with any matrix.
Therefore, 
we may write for $n\ge2$,
\begin{align}
\hspace{-.2em}c^p_\gamma(\bar P^2&+2\bar m \bar P)^n=\nonumber\\
&\bar P\Big[ (\bar P +2\bar m) c^p_\gamma(\bar P^2+2\bar m \bar P)^{n-2}(\bar P +2\bar m)\Big]\bar P\,,
\end{align}
with analogous expressions for $C$, $K$, and $M$, 
as well as $c^p_\gamma$ replaced by $\tilde{k}^p_\gamma$.
This implies that all terms with $n\ge2$ 
in our above expansion of $\Gamma^{(2)}(\bar P)$ 
take the generic form $\bar P [\ldots] \bar P$, 
where the square brackets contain some matrix polynomial in $p^\mu$.
But terms of this form can be ignored for the purpose of extracting the pole structure, 
as is apparent from our discussion 
in the context of Eqs.~(\ref{GenPole}) and~(\ref{InvGenPole}).

The structure of the pole is governed by the remaining terms 
corresponding to $n=0,1$. 
At this point,
the order of various matrices must be determined carefully. 
Consider, 
for example, 
the term 
\begin{equation}\label{n=1_term}
C'(\bar\beta,c^p_p,\tilde{k}^p_p)(\bar P^2+2\bar m \bar P)c^p_\gamma\,,
\end{equation}
which represents the $n=1$ contribution 
to the fourth line in Eq.~(\ref{LVansatzP}).
In this expression, 
$c^p_\gamma$ and $(\bar P^2+2\bar m \bar P)$ 
may still be interchanged freely,
but the pole is determined by terms linear in $\bar P$.
One might then be tempted simply to drop $\bar P^2$ contributions. 
However, 
this yields either
$2\bar m C'(\bar\beta,c^p_p,\tilde{k}^p_p)\bar P c^p_\gamma$ or
$2\bar m C'(\bar\beta,c^p_p,\tilde{k}^p_p)c^p_\gamma\bar P $ 
depending on the choice of matrix ordering 
in the original $n=1$ term~(\ref{n=1_term}). 
Moreover, 
regardless of which one of these two choices is adopted, 
the resulting expression fails to be of the form 
${\cal Z}^{-1}_R \bar P$ required by Eq.~(\ref{InvGenPole}) ,
where ${\cal Z}^{-1}_R$ is a number 
and {\em not} a matrix.

The basic idea behind the selection of the proper ordering 
is as follows.
We parametrize all possible ordering choices, 
arrange this general expression into series in $\bar P$, 
and then fix the parameter 
(and thus the ordering choice) 
such that an expression with the structure demanded by Eq.~(\ref{InvGenPole}) 
emerges. 
Let us demonstrate this idea explicitly for the above sample term~(\ref{n=1_term}), 
where we may ignore 
the $C'(\bar\beta,c^p_p,\tilde{k}^p_p)$ coefficient 
in the present context.
We write 
\begin{eqnarray}\label{ordering_parametrization}
\hspace{-2em}(\bar P^2+2\bar m \bar P)c^p_\gamma &=& 
\zeta(\bar P^2+2\bar m \bar P)c^p_\gamma\nonumber\\
&&{}+(1-\zeta)c^p_\gamma(\bar P^2+2\bar m \bar P)\nonumber\\
&=&2\zeta (2c^p_p\bar P-\bar P c^p_\gamma \bar P)\nonumber\\
&&{}+(1-2\zeta)c^p_\gamma(\bar P^2+2\bar m \bar P)\,,
\end{eqnarray}
where $\zeta$ is a free coefficient parametrizing the matrix ordering. 
To arrive at the last equality, 
we have employed the result 
$\{\bar P^2+2\bar m \bar P,c^p_\gamma\}=4c^p_p \bar P -2\bar P
 c^p_\gamma \bar P$,  
which follows from an explicit evaluation of the anticommutator.
The next step is to fix $\zeta$ by requiring 
compatibility with Eq.~\rf{InvGenPole}.
This is achieved by
choosing $\zeta=\frac{1}{2}$, 
which eliminates the offending terms in the last line. 
We conclude that
the following symmetrized representation of Eq.~(\ref{LVansatzP}), 
\begin{eqnarray}\label{LVansatzPsymm}
\hspace{0em}\Gamma^{(2)}(\bar P) & = &
A(\bar P{}^{2}+2\bar m\bar P+\bar \beta,c^p_p,\tilde{k}^p_p)\bar P\nonumber\\
&&\hspace{-0em}{}+\bar m A(\bar P{}^{2}+2\bar m\bar P+\bar \beta,c^p_p,\tilde{k}^p_p)\openone\nonumber\\
&&\hspace{-0em}{}-M(\bar P{}^{2}+2\bar m\bar P+\bar \beta,c^p_p,\tilde{k}^p_p)\openone\nonumber\\
&&\hspace{-0em}{}+\half\big\{C(\bar P{}^{2}+2\bar m\bar P+\bar \beta),c^p_\gamma\big\}\nonumber\\
&&\hspace{-0em}{}-\half\bar x \big\{A(\bar P{}^{2}+2\bar m\bar P+\bar\beta,c^p_p,\tilde{k}^p_p),c^p_\gamma\big\}\nonumber\\
&&\hspace{-0em}{}+\half\big\{K(\bar P{}^{2}+2\bar m\bar P+\bar \beta),\tilde{k}^p_\gamma\big\}\nonumber\\
&&\hspace{-0em}{}-\half\bar y \big\{A(\bar P{}^{2}+2\bar m\bar P+\bar \beta,c^p_p,\tilde{k}^p_p),\tilde{k}^p_\gamma\big\}\,,\qquad
\end{eqnarray}
possesses the proper matrix ordering 
when an expansion in powers of $\bar P$ 
with the structure~(\ref{InvGenPole}) is desired.

From a practical perspective, 
the above discussion implies
that we may replace
\begin{equation}\label{order_replace}
\half\big\{C(\bar P{}^{2}+2\bar m\bar P+\bar \beta),c^p_\gamma\big\}
\to
C(\bar\beta)c^p_\gamma
+2C'(\bar\beta)c^p_p\bar P
\end{equation}
for the purpose of extracting the pole
with analogous relations for the remaining anticommutators 
in Eq.~(\ref{LVansatzPsymm}).
To leading order in Lorentz violation, 
we find
\begin{eqnarray}\label{LVansatzPsymmExpan}
\hspace{0em}\Gamma^{(2)}(\bar P) & = &
{}+\Big[\bar m A(\bar \beta,c^p_p\tilde{k}^p_p)\openone-M(\bar \beta,c^p_p,\tilde{k}^p_p)\openone\nonumber\\
&&\hspace{2em}{}+C(m_{\rm ph}^2) c^p_\gamma
-\bar x A(m_{\rm ph}^2) c^p_\gamma\nonumber\\
&&\hspace{2em}{}+K(m_{\rm ph}^2) \tilde{k}^p_\gamma
-\bar y A(m_{\rm ph}^2) \tilde{k}^p_\gamma\Big]\nonumber\\
&&\hspace{-0em}{}+\Big[
A(\bar \beta,c^p_p,\tilde{k}^p_p)
+2\bar m^2 A'(\bar \beta,c^p_p,\tilde{k}^p_p)\nonumber\\
&&\hspace{2em}{}-2\bar m M'(\bar \beta,c^p_p,\tilde{k}^p_p)\nonumber\\
&&\hspace{2em}{}-2\bar x A'(m_{\rm ph}^2)c^p_p-2\bar y A'(m_{\rm ph}^2)\tilde{k}^p_p\nonumber\\
&&\hspace{2em}{}+2C'(m_{\rm ph}^2)c^p_p
+2K'(m_{\rm ph}^2)\tilde{k}^p_p
\Big]\bar P\nonumber\\
&&\hspace{-0em}{}+\bar P [\ldots]\bar P\,,
\qquad
\end{eqnarray}
where we have employed the same notation 
as in Eqs.~(\ref{x_eq}) and~(\ref{y_eq}).
The first square bracket needs to vanish 
since $\Gamma^{(2)}(0)=0$ at the pole. 
We thus recover the result~(\ref{LV_pole_eq}). 
The second square bracket 
is identified as ${\cal Z}^{-1}_R$.


\section{Quantization of the Dirac field in the presence of
Lorentz violation}
\label{canonical}

In this appendix, 
we carry out the explicit quantization 
of the Dirac field in the presence of Lorentz violation.
One possible approach is 
to use a field redefinition $\psi = A\chi$ that 
transforms the terms with time derivatives 
to the standard Dirac form~\cite{fermion-field-redefinition,quant1}.
Alternatively,
one may use an unconventional scalar product in spinor space 
to bypass the Hermiticity issues 
associated with unconventional time derivatives~\citep{Kostelecky:2010ze}.
The method presented below is based on this latter approach,
is more direct, 
maintains spinor coordinate covariance, 
is more direct,
maintains spinor coordinate covariance, 
and introduces explicit creation and annihilation
operators for the particle modes corresponding to the physical field $\psi$
as well as an expression for the Hamiltonian in terms of these.
These features are particularly suitable for the quantization of the spinor
equation of motion~(\ref{eom-psi_in}).

We start with the free-fermion Lagrange density
\begin{equation}\label{L}
\mathcal{L}_f=\bar\psi(i\Gamma^\mu \partial_\mu-M)\psi\,,
\end{equation}
with
\begin{align}\label{Gamma-M2}
\Gamma^\mu=&\,\, \gamma^\mu+c^{\mu\nu}\gamma_\nu+d^{\mu\nu}\gamma_5\gamma_\nu
+if^\mu\gamma_5+\frac12 g^{\lambda\nu\mu}\sigma_{\lambda\nu}+e^\mu\,,\\
M=&\, \,m+a^\mu\gamma_\mu+b^\mu\gamma_5\gamma_\mu+{1\over2}H^{\mu\nu}\sigma_{\mu\nu}\,.
\end{align}
The Lorentz-violating background 
is taken as real valued,
so that $\Gamma^\mu{}^\dagger=\gamma^0\Gamma^\mu\gamma^0$ 
and $M^\dagger=\gamma^0M\gamma^0$.
The canonical momentum is given by
\begin{equation}\label{momentum}
\pi=\frac{\partial_R\mathcal{L}_f}{\partial{\dot\psi}}=i\bar\psi\Gamma^0\,,
\end{equation}
and the canonical Hamiltonian becomes 
\begin{align}
H=&\int d^3x\bigl[\pi\dot\psi-\mathcal{L}_f\bigr]=
\int d^3x\,\bar\psi(-i\,\vec{\Gamma}\!\cdot\!\vec{\nabla}+M)\psi\nonumber\\
=&\int d^3x\,\pi(\Gamma^0)^{-1}(-\vec{\Gamma}\!\cdot\!\vec{\nabla}-iM)\psi\,.
\label{can-H}
\end{align}
The matrix $\Gamma^0$ is indeed 
invertible for perturbatively small Lorentz violation~\cite{quant1}.
From this Hamiltonian, 
one can recover the equation of motion
\begin{equation}\label{psi}
\dot\psi(x)=\frac{\delta H}{\delta\pi(x)}=(\Gamma^0)^{-1}(-\vec{\Gamma}\!\cdot\!\vec{\nabla}-iM)\psi\,.
\end{equation}

Next, 
we expand the solutions of the equation of motion for $\psi$ 
in Fourier modes,
\begin{align}
\psi(x)=&\int\!\frac{d^3\vec{p}}{(2\pi)^3}\,\frac{m}{\omega_p}
\sum_{s=1,2}\!\!\big[
b_s(\vec p)u^{s}(\vec p)e^{-ip\cdot x}
\nonumber\\
&
\hspace{10em}+d_s^\dagger(\vec p)v^s(\vec p)e^{ip\cdot x}\big]\,,
\label{fourier-expansion-psi}
\end{align}
where $\omega_p\equiv+(m^2+\vec{p}^{\,2})^{1/2}$
denotes the conventional fermion energies, 
and $p^0$ takes the absolute value of the respective (four) eigenvalues
of $(\Gamma^0)^{-1}(\Gamma^ip^i+M)$.
The latter are, 
in general, 
all different. 
But for small enough Lorentz-violating parameters,
two of them 
(which we will denote $p^0_{u,s}$) are positive,
and two ($-p^0_{v,s}$) are negative~\cite{quant1}.
Below, 
we will show that 
these eigenvalues are, 
in fact, 
real.
The corresponding eigenvectors are 
$u^s(\vec p)$ and $v^s(\vec p)$,
which satisfy
\begin{equation}\label{eq-u,v}
(\Gamma^\mu p_\mu - M)u^s(\vec p)=0\,,\quad(\Gamma^\mu p_\mu + M)v^s(\vec p)=0\,.
\end{equation}
Here, 
Latin superscripts $r,s, \ldots$ from the middle of the alphabet 
label the spin-type state.
To quantize, 
we replace the Poisson brackets with $i\times$anticommutators:
\begin{align}
[\psi_a (\vec x,t),\bar\psi_b (\vec y,t)]_+=&
\;[\psi_a(\vec x,t),-i\pi_c(\vec y,t)]_+(\Gamma^0)^{-1}_{cb}\nonumber\\
=&\;(\Gamma^0)^{-1}_{ab}\delta^3(\vec x-\vec y)
\label{psi-anticomm1}
\end{align}
and
\begin{equation}\label{psi-anticomm2}
[\psi_a(\vec x,t),\psi_b(\vec y,t)]_+=
[\bar\psi_a(\vec x,t),\bar\psi_b(\vec y,t)]_+=0\,,
\end{equation}
where Latin subscripts $a, b, c, \ldots$ from the beginning of the alphabet
denote spinor components.
We now proceed to check that 
the anticommutation relations~(\ref{psi-anticomm1}) 
and (\ref{psi-anticomm2}) 
correspond to taking for the oscillators the following 
nonzero anticommutators:
\begin{equation}
[b_r(\vec p),b_{s}^{\dagger}(\vec q)]_+=
[d_r(\vec p),d_{s}^{\dagger}(\vec q)]_+
=(2\pi)^3\frac{\omega_p}{m}\delta^3(\vec q-\vec p)\,\delta_ {rs}\,.
\label{b,d-comm}
\end{equation}
With these relations, 
we find from Eq.~(\ref{fourier-expansion-psi}),
\begin{align}
\hspace{-.5em}[\psi_a(\vec x,t),\bar\psi_b(\vec y,t)]_+=
&\int \! \frac{d^3\vec{p}}{(2\pi)^3}\,\frac{m}{\omega_p}\sum_s\big[
u_a^s(\vec p)\bar u_b^s(\vec p)\nonumber\\
&
{}+v^s_a(-\vec p)\bar v^s_b(-\vec p)\big]
e^{+i\vec p\cdot(\vec x-\vec y)}\,.
\label{psi-anticomm3}
\end{align}

To see that 
the matrix $\Gamma_0^{-1}(\vec{\Gamma}\!\cdot\!\vec{p}+M)$ 
has real eigenvalues,
we note that
\begin{equation}\label{innerproduct}
\langle u|v\rangle\equiv \bar u \Gamma^0 v
\end{equation}
represents an (unconventional) inner product on spinor space.
In this expression, $u$ and $v$ are arbitrary spinors 
with the usual notation $\bar u\equiv u^\dagger\gamma^0$.
The definition~(\ref{innerproduct}) clearly satisfies 
the appropriate linearity conditions of an inner product.
To establish positive definiteness, 
note first that 
$\bar u \Gamma^0 u$ is always real 
since $(\Gamma^0)^\dagger=\gamma^0\Gamma^0\gamma^0$.
Moreover, 
$\gamma^0\Gamma^0$ 
differs from $\openone$ only by SME corrections,
which implies positive definiteness 
for sufficiently small Lorentz violation~\cite{quant1}.
The matrix $\Gamma_0^{-1}(\vec{\Gamma}\!\cdot\!\vec{p}+M)$ 
turns out to be Hermitian with respect to 
the modified inner product~(\ref{innerproduct}),
which shows that 
this matrix indeed possesses real eigenvalues, 
as claimed.
Moreover, 
the presumed small size of the SME corrections 
implies that 
these eigenvalues are perturbations around the conventional ones, 
so two eigenenergies are positive and two negative~\cite{quant1}. 
The four corresponding eigenspinors
are orthogonal and can be normalized in the usual way:
\begin{equation}\label{norm-u,v}
\bar u^r(\vec p)\Gamma^0 u^s(\vec p)=
\bar v^r(-\vec p)\Gamma^0 v^s(-\vec p)=
\frac{\omega_p}{m}\,\delta_{rs}\,.
\end{equation}
As $\{u^1(\vec p),u^2(\vec p),v^1(-\vec p),v^2(-\vec p)\}$ 
forms a complete set of eigenspinors of the operator 
$(\Gamma^0)^{-1}(\vec{\Gamma}\!\cdot\!\vec{p}+M)$, 
we have the completeness relation
\begin{equation}
\label{is-matrix}
\sum_{s=1}^2\big[u^s_a(\vec p)\bar u^s_c(\vec p)+
v^s_a(-\vec p)\bar v^s_c(-\vec p)\big]\Gamma^0_{cb}
=\frac{\omega_p}{m}\;\delta_{ab}\,.
\end{equation}
We can now use this completeness relation in Eq.~(\ref{psi-anticomm3}),
recovering the result of Eq.~(\ref{psi-anticomm1}).

As a further application, 
let us derive an expression for the Hamiltonian
in terms of the oscillators. 
The quantum-field version of Eq.~(\ref{can-H}) is given by 
\begin{equation}\label{can-H2}
H=\int \! d^3x\, :\bar\psi(-i\,\vec{\Gamma}\!\cdot\!\vec{\nabla}+M)\psi:\,.
\end{equation}
With the Fourier decomposition~(\ref{fourier-expansion-psi}) we find,
using the equations of motion~(\ref{eq-u,v}) 
for $u^s(\vec p)$ and $v^s(\vec p)$, 
an explicitly positive-definite expression for the Hamiltonian,
\begin{equation}\label{can-H3}
H=\int\! \frac{d^3\vec{p}}{(2\pi)^3}\,\frac{m}{\omega_p}
\sum_{s=1}^2\big[
E^s_u b_s^\dagger(\vec p)b_s(\vec p)+
E^s_v d_s^\dagger(\vec p)d_s(\vec p)\big]\,,
\end{equation}
where $E^s_u$ and $E^s_v$ are the (positive) energies 
corresponding to the solutions of Eq.~(\ref{eq-u,v}).
We also note the useful expressions
\begin{eqnarray}\label{oscillator-expression-b}
b_s^\dagger(\vec p)&=&
\int\! d^3x\,e^{-i p\cdot x}\bar\psi(x)\Gamma^0u^s(\vec p)\,,\\
d_s^\dagger(\vec p)&=&
\int\! d^3x\,e^{-i p\cdot x}\bar v_s(\vec p)\Gamma^0\psi(x)\,,
\label{oscillator-expression-d}
\end{eqnarray}
and their Hermitian conjugates.

The time-ordered product is defined in the usual way~\cite{Tprod} 
and satisfies
\begin{equation}
T\psi_a(x)\bar\psi_b(y)=
\langle0|T\psi_a(x)\bar\psi_b(y)|0\rangle
+:\psi_a(x)\bar\psi_b(y):\,,
\end{equation}
where
\begin{equation}\label{Tfeynman}
\langle0|T\psi_a(x)\bar\psi_b(y)|0\rangle\equiv iS(x-y)_{ab}
\end{equation}
is the modified Feynman propagator:
\begin{equation}\label{FeynExpr}
S(x-y)=\int\frac{d^4k}{(2\pi)^4}
\frac{e^{-ik\cdot(x-y)}}{\Gamma^\mu k_\mu-M+i\epsilon}\,.
\end{equation}
Indeed, 
one can verify that
\begin{equation}\label{FeynCheckExpr}
(i\Gamma^\mu\partial_\mu-M)\langle0|T\psi(x)\bar\psi(y)|0\rangle=i\delta^4(x-y)\,.
\end{equation}
Here, 
we used that 
$\psi(x)$ satisfies the modified Dirac equation~(\ref{psi})
and the anticommutator relation~(\ref{psi-anticomm1}).


\section{Canonical quantization of the radiation field with
Lorentz-violating parameter $\bm{\tilde k^{\mu\nu}}$}
\label{canonical-radiation}

Our starting point is the following Lorentz-violating Stueckelberg
Lagrange density in $\xi=1$ Feynman gauge,
\begin{equation}
\mathcal{L}_\gamma=
-\tfrac{1}{4} \tilde\eta^{\alpha\beta}\tilde\eta^{\mu\nu}
F_{\alpha\mu} F_{\beta\nu}
-\tfrac{1}{2}(\partial_\mu\tilde\eta^{\mu\nu}\!A_\nu)^2
+\tfrac{1}{2} m_\gamma^2 A_\mu\tilde\eta^{\mu\nu}\!A_\nu\,,
\label{radiation-Lagrangian}
\end{equation}
where $\tilde\eta^{\mu\nu}=\eta^{\mu\nu}+\tilde k^{\mu\nu}$, 
with $\tilde k^{\mu\nu}=\tilde k^{\nu\mu}$ and $\tilde k^\mu{}_\mu=0$.
The quantization of such photon models 
has recently been studied by various authors~\citep{quant3}.
Here, 
we summarize the main results 
and tailor the presentation to the case at hand.

We begin by finding the canonical momenta
\begin{equation}
\Pi^\mu=\frac{\partial\mathcal{L}}{\partial(\partial_0A_\mu)}=
\tilde\eta^{\mu\alpha}\tilde\eta^{0\beta}F_{\alpha\beta}
-\tilde\eta^{0\mu}(\partial_\mu\tilde\eta^{\mu\nu}A_\nu)\,,
\label{PiA}
\end{equation}
and impose the fundamental equal-time commutation relations
\begin{align}
\label{comrel-AA-PiPi}
[A_\mu(t,\vec x),A_\nu(t,\vec y)]
&=[\Pi^\mu(t,\vec x),\Pi^\nu(t,\vec y)]=0\,, \\
[A_\mu(t,\vec x),\Pi^\nu(t,\vec y)]
&=i\delta_\mu^\nu\delta^3(\vec x-\vec y)\,.
\label{comrel-APi}
\end{align}
From Eq.~(\ref{comrel-AA-PiPi}) it follows that 
the spatial derivatives of $A_\mu$ commute at equal times.
Using Eqs.~(\ref{comrel-APi}) and (\ref{PiA}), 
one then deduces that
\begin{equation}
[\dot A_\mu(t,\vec x),A_\nu(t,\vec y)]=
i(\tilde\eta^{00})^{-1}\bar\eta_{\mu\nu}\delta^3(\vec x-\vec y)\,,
\label{comrel-dotAA}
\end{equation}
where $\bar\eta_{\mu\nu}$ is defined as the inverse of 
$\tilde\eta^{\mu\nu}$,
\begin{equation}
\tilde\eta^{\mu\alpha}\bar\eta_{\alpha\nu}=\delta^\mu_\nu\,,
\qquad \bar\eta_{\mu\nu}\approx\eta_{\mu\nu}-\tilde k_{\mu\nu}\,.
\end{equation}
The equation of motion following from (\ref{radiation-Lagrangian})
is
\begin{equation}
\left(\partial_\alpha \tilde\eta^{\alpha\beta}\partial_\beta
+m_\gamma^2\right) A_\mu =0\,,
\label{eom-A}
\end{equation}
which implies that 
the dispersion relation is the same for all four modes.
In other words, 
our model is strictly free of any birefringence.

Consider now the vacuum expectation value of the time-ordered product
\begin{equation}
\langle0|T\,A_\mu(x)A_\nu(y)|0\rangle\,.
\end{equation}
Acting on it with the kinetic operator we find
\begin{align}
&\left(\partial_\alpha \tilde\eta^{\alpha\beta}\partial_\beta
+m_\gamma^2\right)_x\langle0|T\,A_\mu(x)A_\nu(y)|0\rangle=
\nonumber\\
&\quad=
\langle0|T\left[(\partial_\alpha \tilde\eta^{\alpha\beta}\partial_\beta
+m_\gamma^2)A_\mu(x)\right]A_\nu(y)|0\rangle\nonumber\\
&\qquad\qquad{}+\delta(x_0+y_0)\langle0|[\tilde\eta^{0\beta}\partial_\beta
A_\mu(x)A_\nu(y)]|0\rangle\nonumber\\
&\quad= \delta(x^0-y^0)\tilde\eta^{00}\langle0|[\dot A_\mu(x),A_\nu(y)]|0\rangle
\nonumber\\
&\quad=i\delta^4(x-y) \bar \eta_{\mu \nu}\,,
\end{align}
where we have used relation~(\ref{comrel-dotAA}).
We infer that 
$\langle0|T\,A_\mu(x)A_\nu(y)|0\rangle$ 
must indeed be equal to the (modified) Feynman propagator:
\begin{equation}
\langle0|T\,A_\mu(x)A_\nu(y)|0\rangle
=-i\!\!\int \!\! \frac{d^4k}{(2\pi)^4}
\frac{e^{-ik\cdot(x-y)}}{k_\alpha\tilde\eta^{\alpha\beta}k_\beta-
m_\gamma^2+i\epsilon}\bar\eta_{\mu\nu}\,,
\label{radiation-feynman-propagator1}
\end{equation}
paralleling the Lorentz-invariant case.

It is useful to cast Eq.~(\ref{radiation-feynman-propagator1}) 
in an alternative form.
To this end, we write
\begin{widetext}
\begin{equation}
\frac{1}{k_\alpha\tilde\eta^{\alpha\beta}k_\beta-
m_\gamma^2+i\epsilon}
=\frac{(\tilde\eta^{00})^{-1}}{k_{0+}+k_{0-}}
\left(\frac{1}{k_0-k_{0+}+i\epsilon}-\frac{1}{k_0+k_{0-}-i\epsilon}\right)\,,
\label{denominator-rel}
\end{equation}
where $\pm k_{0\pm}(\vec k)$ are the two roots of the dispersion relation
\begin{equation}
k_\alpha\tilde\eta^{\alpha\beta}k_\beta - m_\gamma^2=0\,,
\label{radiation-dispersion-relation}
\end{equation}
which follows from Eq.~(\ref{eom-A}).
Here, 
$k_{0\pm}(\vec k)$ are both taken positive, 
so that the roots of the dispersion relation have alternate signs,
as in the Lorentz-invariant case.
(This is justified,
for example, 
in concordant frames,
in which we may take $|\tilde k^{\mu\nu}|\ll 1$ 
on experimental grounds.)
Note that, generally, $k_{0+}(\vec k)\ne k_{0-}(\vec k)$, but
\begin{equation}
k_{0\pm}(\vec k)= k_{0\mp}(-\vec k)
\end{equation}
follows because Eq.~(\ref{radiation-dispersion-relation}) 
is even in the components of the momentum.
Using Eq.~(\ref{denominator-rel}), 
one derives easily that
the Feynman propagator (\ref{radiation-feynman-propagator1}) 
can be represented as
\begin{align}
\langle0|T\,A_\mu(x)A_\nu(y)|0\rangle&=-\bar\eta_{\mu\nu} (\tilde\eta^{00})^{-1}
\int\frac{d^3\vec{k}}{(2\pi)^3}\frac{1}{k_{0+}+k_{0-}}
\left[\theta(x^0-y^0)e^{-i\left(k_{0+}\,x_0-\vec k\cdot\vec x\right)}+
\theta(y^0-x^0)e^{i\left(k_{0+}\,x_0-\vec k\cdot\vec x\right)}\right].
\label{radiation-feynman-propagator2}
\end{align}
Note that 
in both terms of Eq.~(\ref{radiation-feynman-propagator2}) 
only the positive
root for $k_0$ appears in the exponentials.
It is also worthwhile pointing out the factor $k_{0+}+k_{0-}$
(which, 
unlike the individual roots $k_{0\pm}$, 
is an even function of $\vec k$) 
that appears in the denominator,
replacing the usual factor $2k_0$.

Let us now try to represent the dynamical system
described above by the simple mode expansion
\begin{equation}
A_\mu(x)=\int\! \frac{d^3\vec{k}}{(2\pi)^3N(k)} \sum_{\lambda=0}^3
\big[a^{(\lambda)}(k)
\epsilon_\mu^{(\lambda)}(k)e^{-ik\cdot x}+a^{(\lambda)\dagger}(k)
\epsilon_\mu^{(\lambda)*}(k)e^{ik\cdot x}\big]\,,
\label{A-mode-expansion}
\end{equation}
where $k^\mu=(k_{0+},\vec{k})$ 
satisfies the dispersion relation~(\ref{radiation-dispersion-relation}),
and $N(k)$ is a (yet to be determined) function.
Next,
we posit creation and annihilation operators 
satisfying the nonzero commutation relations
\begin{equation}
[a^{(\lambda)}_\mu(k),a^{(\lambda')\dagger}_\nu(k')]=
-(2\pi)^3\eta^{\lambda\lambda'}M(\vec k)\delta^3(\vec k-\vec k')\,.
\label{comrel-oscillators}
\end{equation}
Here, 
the normalization $M(\vec{k})$ 
is to be chosen later.
With Eqs.~(\ref{A-mode-expansion}) and~(\ref{comrel-oscillators}) at hand,
the time-ordered product $\langle0|T\,A_\mu(x)A_\nu(y)|0\rangle$
can be expressed as
\begin{align}
\langle0|T\,A_\mu(x)A_\nu(y)|0\rangle&=-\int\frac{d^3\vec{k}}{(2\pi)^3}
\frac{M(k)}{N(k)^2}\sum_{\lambda,\lambda'}\epsilon_\mu^{(\lambda)}(k)
\epsilon_\nu^{(\lambda')}(k)\eta_{\lambda\lambda'}
\left[\theta(x^0-y^0)e^{-ik\cdot(x-y)}+\theta(y^0-x^0)e^{ik\cdot(x-y)}\right].
\label{A-propagator-expansion}
\end{align}
\end{widetext}
Comparing Eq.~(\ref{radiation-feynman-propagator2}) 
with our earlier form of 
the Feynman propagator~(\ref{A-propagator-expansion}),
we deduce
\begin{equation}
\frac{M(k)}{N(k)^2}\sum_{\lambda,\lambda'}\epsilon_\mu^{(\lambda)}(k)
\epsilon_\nu^{(\lambda')}(k)\,\eta_{\lambda\lambda'}=
\frac{1}{k_{0+}+k_{0-}}(\tilde\eta^{00})^{-1}\bar\eta_{\mu\nu}\,.
\label{condition-constants}
\end{equation}
While there are various ways to satisfy Eq.~(\ref{condition-constants}),
we will take
\begin{align}
\label{constants-N,M}
N(k)&=M(k)=k_{0+}+k_{0-}\\
\sum_{\lambda,\lambda'}
\epsilon_\mu^{(\lambda)}(k)\epsilon_\nu^{(\lambda')}(k)\, 
\eta_{\lambda\lambda'}&= (\tilde\eta^{00})^{-1}\bar\eta_{\mu\nu}
\label{polarization-sum}
\end{align}
in what follows.
It is a nontrivial but straightforward exercise 
to verify that 
with the choices (\ref{constants-N,M}) and (\ref{polarization-sum}) 
the equal-time commutators (\ref{comrel-AA-PiPi}) and (\ref{comrel-APi}) 
are correctly represented.
As an aside, 
we note that 
the usual relation
$[\dot A_\mu(t,\vec x),\dot A_\nu(t,\vec y)]=0$
is no longer valid.

Equation~(\ref{polarization-sum}) implies the normalization condition
\begin{equation}
\epsilon^{(\lambda)}_\mu(k)\epsilon^{(\lambda')}_\nu(k)\,\tilde\eta^{\mu\nu}
=(\tilde\eta^{00})^{-1}\eta^{\lambda\lambda'}\,.
\label{normalization-polarization-vecs}
\end{equation}
It is convenient to select the timelike, unphysical polarization mode as
\begin{equation}
\epsilon^{(0)}_\mu(k)=\frac{k_\mu}{m_\gamma\sqrt{\tilde\eta^{00}}}\,.
\label{epsilon0}
\end{equation}
in accordance with the spin sum~(\ref{normalization-polarization-vecs}).
The three transverse, physical polarization modes
$\epsilon^{(\lambda)}_\mu(k)$ for $\lambda=1,2,3$ are orthonormal,
spacelike vectors, orthogonal to $k^\mu$,
with respect to the effective metric $\tilde\eta^{\mu\nu}$.
This choice of the transverse modes corresponds to defining the
physical states satisfying
\begin{equation}
\langle\mbox{phys}|\partial_\mu \tilde\eta^{\mu\nu}A_\nu|\mbox{phys}\rangle=0\,.
\end{equation}
It follows from Eqs.~(\ref{polarization-sum}) and~(\ref{epsilon0}) that
the physical-polarization sum becomes
\begin{equation}
\tilde\eta^{00}\sum_{\lambda=1}^3
\epsilon_\mu^{(\lambda)}(k)\epsilon_\nu^{(\lambda)}(k)=
-\bar\eta_{\mu\nu}+\frac{k_\mu k_\nu}{m_\gamma^2}.
\label{polarization-sum2}
\end{equation}

Finally, we note that
with the choice~(\ref{constants-N,M}) 
the 3-momentum measure 
appearing in Eq.~(\ref{A-mode-expansion}) 
satisfies the property
\begin{align}
&\int\frac{d^3k}{k_{0+}+k_{0-}}f(k_{0+},\vec k)=
\nonumber\\
&\quad{}=\frac{\tilde\eta^{00}}{2}
\int d^4k\,\delta^4(k_\alpha\tilde\eta^{\alpha\beta}k_\beta-m_\gamma^2)f(k_0,\vec k)\,,
\label{measure-property}
\end{align}
where $f(k^\mu)$ is an arbitrary even function of $k^\mu$.



\begin{thebibliography}{99}
\bibitem{ghostfn}
Strictly speaking,
we should include the action for the ghost fields~\cite{Stueckelberg}.
As in this paper we will be mainly interested in external physical states,
we will disregard them in this work. 
Note also that 
we have switched to the usual non-Hermitian representation of the fermion Lagrangian.
\bibitem{footnote-1}
The Feynman rules for the insertion in Fig.~1 
and for the photon propagator in Fig.~2 
are irrelevant at this stage.
\bibitem{footnote1}
Beyond linear order in Lorentz violation 
and when additional interactions are considered 
(e.g., parity-violating electroweak physics)
the general structure of $\delta\Sigma$ becomes 
more intricate than that given in Eq.~\rf{deltaSigma}.
\bibitem{PertCom}
The standard premise in perturbation theory posits that
tree-level single-particle states 
and those in the fully interacting theory
are in one-to-one correspondence 
and can differ at most by small corrections. 
In the present situation of small couplings $\alpha$ and $c^{\mu\nu}$ 
and with properly treated perturbation-theory infinities,
there seems to be no reason to abandon 
this premise.
\bibitem{ZCom}
This is not strictly necessary if the spinor normalizations 
are adjusted accordingly.
However, 
the $\slashed p$ kinetic operator is the only term in $P_0+\delta P'$
independent of the model's parameters 
such as $m$ and $\alpha$ in conventional QED
or $m$, $\alpha$, $c^{\mu\nu}$, and $\tilde{k}^{\mu\nu}$ 
in the present case.
It follows that 
maintaining the tree-level normalization of the $\slashed p$ contribution 
permits,
e.g.,
an unambiguous identification 
of the corrections to $m$ and $c^{\mu\nu}$.
\bibitem{LIcomment}
This is the most general expression 
compatible with both Lorentz and parity symmetry.
\bibitem{PropComment}
If $c^p_\gamma$ and $\tilde{k}^p_\gamma$ happen to be proportional,
we may select another rescaling of the coordinates that
distributes the Lorentz-violating effects between the electron and the photon in such a way that
$c^p_\gamma\to c'{}^p_\gamma$ and $\tilde{k}^p_\gamma\to\tilde{k}'{}^p_\gamma$
with $c'{}^p_\gamma$ and $\tilde{k}'{}^p_\gamma$ 
linearly independent.
\bibitem{k_comment}
Here, 
we have anticipated the loop-corrected version
$\tilde k_\phy^{\mu\nu}$ of the coefficient $\tilde k^{\mu\nu}$.
While we do not compute
its value in this work,
it will differ from $\tilde k^{\mu\nu}$ by 
an expression of order $\alpha$,
so that it can be substituted for the latter in Eq.~(\ref{P-phys}),
as we disregard terms of order $\alpha^2$.
\bibitem{c=const_comment}
Note that our $c$ coefficient is spacetime constant,
so that with our interpretation
the spin connection vanishes.
Note also that $c$ is traceless, 
so $e \equiv \det e^{\mu}_a = 1+{\cal O}(c^2)$.
\bibitem{Tprod}
In the conventional case,
the time-ordered product is coordinate independent 
because the Dirac field anticommutes outside the lightcone.
In the present context involving Lorentz violation,
certain SME-parameter combinations 
allow the anticommutator function to be supported also 
in small spacelike regions in the close vicinity of the light cone. 
Extremely large boosts of the coordinate system 
may then spoil coordinate independence of the time-ordered product.
\end{thebibliography}
\end{document}